\documentclass[10pt,journal,twoside]{IEEEtran}
\usepackage{graphicx}

\usepackage{amsmath}
\usepackage{amsfonts}
\usepackage{amssymb}
\usepackage{psfrag}
\usepackage{cite}
\usepackage{xcolor}
\usepackage{multirow}
\usepackage{subfigure}
\usepackage[linesnumbered,lined,commentsnumbered]{algorithm2e}
\usepackage{hyperref}
\newtheorem{theorem}{\underline{Theorem}}
\newtheorem{lemma}{\underline{Lemma}}
\newtheorem{corollary}{\underline{Corollary}}

\newtheorem{remark}{\underline{Remark}}

\newtheorem{proposition}{\underline{Proposition}}

\begin{document}

\setcounter{page}{1}
\newcommand{\half}%
{\raisebox{2.5pt}{\scriptsize 1}{\small
/}\raisebox{-1pt}{\scriptsize 2}}
\newcommand{\greatH}{\stackrel{\textstyle{H_{1}}}{>}}
\newcommand{\lessH}{\stackrel{\textstyle{<}}{H_{0}}}
\newcommand{\great}{\stackrel{\textstyle{\greatH}}{\lessH}}
\newcommand{\greatless}{\raisebox{-.45cm}{$\great$}}
\newcommand{\beq}{\begin{equation}}
\newcommand{\eeq}{\end{equation}}
\newcommand{\EE}{{\rm E}}
\newcommand{\PP}{{\rm P}}
\newcommand{\PPr}{{\rm Pr}}
\newcommand{\TT}{{\rm T}}

\def\ruleone{\rule[-1.5ex]{0cm}{4.0ex}}
\title{Average SEP-Optimal Precoding for Correlated Massive MIMO with ZF Detection: \\An Asymptotic Analysis}
\author{Zheng Dong, Jian-Kang Zhang and He Chen
\thanks{Zheng Dong was with the Department of Electrical and Computer Engineering, McMaster University, Hamilton, ON, Canada, and he is now with the School of Electrical and Information Engineering, The University of Sydney, NSW 2006, Australia (Email: zheng.dong@sydney.edu.au). Jian-Kang Zhang is with the Department of Electrical and Computer Engineering, McMaster University, Hamilton, ON, Canada (Email:jkzhang@mail.ece.mcmaster.ca). He Chen is with the School of Electrical and Information Engineering, The University of Sydney, NSW, Australia (Email:he.chen@sydney.edu.au).
This paper was presented in part at the IEEE International Conference on Communications (ICC), London, UK from 8-12, June, 2015\cite{dong15ICC}.}}
\maketitle
\vspace{-50pt}
\begin{abstract}
This paper investigates the symbol error probability~(SEP) of point-to-point massive multiple-input multiple-output (MIMO) systems using equally likely PAM, PSK, and square QAM signallings in the presence of transmitter correlation. The receiver has perfect knowledge of the channel coefficients, while the transmitter only knows first- and second-order channel statistics. With a zero-forcing~(ZF) detector implemented at the receiver side, we design and derive closed-form expressions of the optimal precoders at the transmitter that minimizes the average SEP over channel statistics for various modulation schemes. We then unveil some nice structures on the resulting minimum average SEP expressions, which naturally motivate us to explore the use of two useful mathematical tools to systematically study their asymptotic behaviors. The first tool is the Szeg\"o's theorem on large Hermitian Toeplitz matrices and the second tool is the well-known limit: $\lim_{x\to\infty}(1+1/x)^x=e$. The application of these two tools enables us to attain very simple expressions of the SEP limits as the number of the transmitter antennas goes to infinity. A major advantage of our asymptotic analysis is that the asymptotic SEP converges to the true SEP when the number of antennas is moderately large. As such, the obtained expressions can serve as effective SEP approximations for massive MIMO systems even when the number of antennas is not very large. For the widely used exponential correlation model, we derive closed-form expressions for the SEP limits of both optimally precoded and uniformly precoded systems. Extensive simulations are provided to demonstrate the effectiveness of our asymptotic analysis and compare the performance limit of optimally precoded and uniformly precoded systems.
\end{abstract}

\section{Introduction}
With recent advances in radio frequency~(RF) chains and integrated circuit designs, massive multiple-input multiple-output~(MIMO)~\cite{Marzetta10, Larsson14feb, Lu14Oct,Wong17book,Jing18TCOM} has emerged as a key enabler to fulfill the unprecedented requirement of the upcoming fifth generation (5G) cellular systems on three generic services, including extreme mobile broadband (eMBB), massive machine-type communications (mMTC), and ultra-reliable and low-latency communications (URLLC)~\cite{Andrews14, Popovski14, Popovski16ComMag}.
In general, massive MIMO technology can be categorized into the point-to-point massive MIMO and the multiuser massive MIMO~\cite{Wu17twc, Lu14Oct}. In this paper, we consider a point-to-point massive MIMO where the receiver has an excess number of antennas than the transmitter ($N_r - N_t\gg 1$). This model depicts a typical scenario where a multi-antenna user transmits to its serving base station (BS) equipped with a large number of antennas in the uplink. To implement the aforementioned massive MIMO system  in a real environment, several significant issues should be addressed. The first one is the drastically increased detection complexity that can restrict the widespread deployment of an extreme large array. Despite the fact the maximum likelihood (ML) detector is universally optimal for the uniformly distributed input, its complexity is prohibitively high even for a moderately sized array. Therefore, for the sake of practicality, we consider the use of a linear zero-forcing (ZF) detector at the receiver side, which has been proved to have near-optimal performance in terms of throughput when the BS array size is large~\cite{Larsson14feb, Kit14jstsp, Lu14Oct}.

Another major limiting factor arising in the massive MIMO fading channels is the acquisition of channel state information (CSI). In fact, the performance of massive MIMO relies critically on the availability of the channel knowledge on both sides in order to harvest the array gain and/or multiplexing gain for increased energy efficiency and data rates. In this paper, we assume that full CSI is available at the receiver side while only first- and second-order channel statistics are available at the transmitter as in\,\cite{JafarICC01}. This reason is that the CSI at the receiver can be obtained by letting the transmitter send training pilots, the number of which is proportional to the number of transmitter antennas\,\cite{Marzetta06Asilomar}. It is worth noting that, for practical massive MIMO systems, the CSI can be erroneous due to the difficulty of channel estimation caused by the large number of transmitter and/or receiver antennas and the potential low transmission power. It has been shown that the CSI imperfection will result in lower nominal received signal-to-noise ratio (SNR) in conventional MIMO systems in~\cite{Murch2007twc}, which has recently been further verified for massive MIMO cases in~\cite{lichengSPL2018}. Since the SNR expressions have the same structure no matter whether the CSI is perfect or not, our analysis and the resultant precoding design can also be applied to the cases with imperfect CSI, which is omitted due to space limitation. On the other hand, the instantaneous CSI at the transmitter is arguably intractable to attain by either employing direct training methods relying on the time division duplex (TDD) channel reciprocity or by using quantized channel feedback in frequency-division duplext (FDD) mode, given the large number of receiver antennas in massive MIMO systems\,\cite{Larsson14feb}. To make massive MIMO systems more feasible, more practical constraints need to be considered, such as channel estimation error, pilot contamination, and transceiver hardware impairments. Therefore, proper pilot design and power control as well as advanced linear precoding or two-layer decoding should be carried out to alleviate these limiting factors. However, these are beyond the scope of this paper, and we refer to interested readers to the relevant work~\cite{Elijah16, Larsson18twc,Hoydis18twc, Emil18arXiv, Debbah14tit} and references therein.

More importantly, the effects of fading correlation in massive MIMO systems also need to be better understood. In practice, due to the size limitation, the antennas at the transmitter side (i.e., user device) can hardly be well separated and hence tend to be correlated~\cite{Murch06, JafarICC01, Nossek03, Paulraj02isit, Paulraj02espc, Narasimhan06}. This kind of channel correlation model is a typical case which can be verified by using a `one-ring' scattering model as considered in~\cite{Foschini00mar}. For example, when the user device is unobstructed and the BS is surrounded by local scatterers. To show the effect of fading correlation more explicitly, we specifically consider an exponential correlation model~\cite{KMS53, Turin62, Loyka01, Veeravalli01, Mestre03, Win08, Gesbert11, Zoltowski14}, which is a simplified one-parameter model in practical environment that can capture the main phenomenon of spatial correlation between antennas, especially for a uniform linear array (ULA). To better understand the channel correlation in such massive MIMO fading channels, we consider both the precoder design and the corresponding performance analysis.

It has been shown that, for a general MIMO system with the first- and second-order channel statistics available at the transmitter, linear precoding is an efficient and effective scheme to significantly alleviate the performance loss caused by the channel fading correlation~\cite{kiessling04, JafarTWC04, moustakas2000, liu09sept}. Previously, the performance of a conventional MIMO system with transmitter correlation and the ZF receiver was first investigated in~\cite{Paulraj02isit}, where it was showed that the resulting signal-to-noise ratio (SNR) of each data stream follows a Chi-squared distribution. Then, the capacity analysis for MIMO systems with the receiver having full CSI and the transmitter having only second-order channel statistical information was investigated in~\cite{JafarICC01}. The impact of transmitter correlation in MIMO systems, with full or average CSI was investigated in~\cite{Nossek03}. The antenna selection strategy with transmitter correlation was studied in~\cite{Narasimhan06}.  However, all the results mentioned above were aimed for the conventional MIMO where the number of antennas at the transmitter and receiver are relatively small (e.g., several tens of antennas). To the best knowledge of the authors, this is the first effort towards the asymptotic symbol error probability~(SEP) analysis of massive MIMO systems with transmitter correlation and channel statistics-based precoding. The main purpose of the asymptotic analysis is to unveil the dominating factors on the average error performance of the considered massive MIMO system when the antenna array size is scaled up. The asymptotic error expression also serves as a useful system design guideline even when the antenna array is moderately large. The potential limitation is that the asymptomatic results may diverge from the exact values for small or moderate MIMO systems.  We will show that for our asymptotic analysis, this limitation is rather mild because the asymptotic SEP converges to the true SEP with tens of transmitter antennas.

The main contributions of this paper are summarized as follows:
\begin{enumerate}
\item We first derive the average SEP of the considered massive MIMO system with a ZF detector as a function of the precoding matrix and the channel covariance matrix for PAM, PSK, and square QAM modulations, respectively. Based on the average SEP expressions, we then construct the optimal precoders in closed-form formats that minimize the average SEP of different modulation schemes. Our approach to obtain the optimal precoder is universal, which depends only on the convexity and monotonic of the average SEP expressions as well as the ZF equalizer itself.
\item  We identify some nice structures of the minimal average SEP achieved by the optimal precoders, which naturally lead us to explore the use of two useful mathematical tools for the systematic study of asymptotic behaviors on their error performance. The first tool is the Szeg\"o's theorem on large Hermitian Toeplitz matrices and the second tool is the well-known limit:  $\lim_{x\to\infty}(1+1/x)^x=e$. By applying these two tools, we obtain neat expressions for the average SEP limits of PAM, PSK and square QAM constellations, which are shown to have a fast convergence speed as the number of the transmitter antennas goes to infinity.  Our result can greatly simplify the error performance evaluation for massive MIMO systems.
\item When the channel covariance matrix is the commonly-used non-symmetric Kac-Murdock-Szeg\"o (KMS) matrix, we derive closed-form expressions of the SEP limits for both optimally precoded and uniformly precoded systems. Moreover, a tight approximation to the distribution of the individual SNR of each sub-channel is also developed for the large array, which explicitly reveals the channel hardening phenomenon of the correlated massive MIMO systems. Specifically, for given transmitted power and noise level, the SNR of each sub-channel converges to some constant value almost surely for each channel realization when the array size is scaled up.
\end{enumerate}

{\bf{Notations}}:
Matrices and column vectors are denoted by boldface characters with uppercase  (e.g., ${\mathbf A}$) and lowercase (e.g., ${\mathbf b}$), respectively. $\otimes$ denotes the Kronecker product.
The $(m, n)$-th entry of ${\mathbf A}$ is denoted by $[{\mathbf A}]_{m,n}$. The $m$-th entry of ${\mathbf b}$ is denoted by $b_m$.
$\left\|\mathbf b\right\|$ denotes the Euclidean norm of $\mathbf b$.
Notation ${\mathbf A}^+$ stands for the pseudo-inverse of ${\mathbf A}$.
The transpose of ${\mathbf A}$ is denoted by ${\mathbf A}^T$.
The Hermitian transpose of ${\mathbf A}$ (i.e., the conjugate and transpose of ${\mathbf A}$) is denoted by
${\mathbf A}^H$.

\section{Precoded Transmission Model with Zero-Forcing Detection \label{sec2}}
\subsection{System Model}
Consider the complex baseband-equivalent model of a narrow-band massive MIMO communication system, with $N_t$ transmitting antennas and $N_r$ receiving antennas. The receiver is assumed to has an excess number of antennas than the transmitter ($N_r - N_t\gg 1$).
The serially transmitted symbols is first
de-multiplexed into a vector signal ${\mathbf s}=[s_1, s_2,
\cdots, s_{N_t}]^T$, and then this vector signal ${\mathbf s}$ is
transformed by an $N_t\times N_t$ full-rank square precoding matrix ${\mathbf F}$
into another vector signal ${\mathbf x}=[x_1, x_2, \cdots,
x_{N_t}]^T={\mathbf F}{\mathbf s}$. Then, each element $x_k$ of
${\mathbf x}$ is fed to the $k$-th transmitter antenna for
transmission. In the space-time communication, $N_t$ transmitting antennas
each can transmit uncoded $M$-ary PAM, PSK or QAM symbols using the same
waveform during the same time interval, but we assume that the
same constellation is used for each antenna. At the receiver array, the discrete received
$N_r\times 1$ signal vector ${\mathbf r}$ can be written as
\begin{eqnarray}\label{model}
{\mathbf r}={\mathbf H}{\mathbf F}{\mathbf s}+{\boldsymbol \xi},
\end{eqnarray}
where ${\mathbf H}$ is the $N_r\times N_t$ complex channel matrix and
${\boldsymbol\xi}$ is the $N_r\times 1$ additive complex noise vector.
Each of the $h_{mn} = [{\mathbf H} ]_{mn}$ represents the subchannel connecting the $n$-th transmitter antenna
to the $m$-th receiver antenna. We also let $\mathbf h_m^T=[h_{m1} ,  \cdots , h_{m N_t} ]$ denote the $m$-th row of ${\mathbf H}$. We assume that only the channels at the transmitter side are correlated while the channels at the receiver side are uncorrelated~\cite{JafarICC01, Paulraj02isit}, since the transmitter antennas of mobile users can hardly be well separated and tend to be correlated while the BS can have a large space where the correlation can be made to be negligible. That is, we assume
\begin{eqnarray}\label{eq:ChannelCorr}
    \mathbb{E}[(\mathbf h_\ell^T)^H \mathbf h_m^T] = \left\{  \begin{array}{ll}
    {\boldsymbol\Sigma}
    & \ell=m, \\ {\mathbf 0} &\ell \neq m.
    \end{array} \right .
\end{eqnarray}

The second-order channel statistic matrix $\mathbf{\Sigma}$ is mainly affected by the large-scale fading, which changes much more slowly than the instantaneous channel coefficients. In practice, $\mathbf{\Sigma}$ can be tracked by the base station with a desired accuracy and is assumed to be known as in~\cite{Caire13tit}.
Throughout this paper, we adopt the following assumptions:
\begin{enumerate}
 \item \textit{The perfect channel estimates are available at the receiver to allow coherent detection, while the transmitter only knows first- and second-order channel statistics $\mathbf{\Sigma}$\cite{JafarICC01}};
 \item \textit{The channel ${\mathbf H}$ is complex Gaussian distributed,
 with zero-mean, and covariance matrix ${\mathbf I}\otimes{\boldsymbol\Sigma}$};
 \item \textit{${\boldsymbol\xi}$ is circularly-symmetric complex Gaussian
 noise with covariance $\sigma^2 {\mathbf I}$};
 \item \textit{Each element of ${\mathbf s}$ is independently and equally likely chosen from PAM, PSK or QAM constellations of the same
size with the covariance matrix of ${\mathbf s}$ being
${\mathbf I}$\cite{Palomar03}};
\item  \textit{The total power budget of the transmitter array is unified to one, and as a consequence the system SNR is defined as $\eta \triangleq 1/\sigma^2$}.
\end{enumerate}
\subsection{Zero-Forcing Equalization}
Suppose that we use zero-forcing equalization to recover the
information symbols. To this end, first we obtain the pseudo
inverse of super-channel matrix ${\mathbf H}{\mathbf F}$, i.e.
\begin{eqnarray}\label{inverse}
({\mathbf H}{\mathbf F})^{+}=\left({\mathbf F}^H{\mathbf
H}^H{\mathbf H}{\mathbf F}\right)^{-1}{\mathbf F}^H{\mathbf H}^H.
\end{eqnarray}

Here we need to explain why the inverse in~(\ref{inverse}) exists.
Under Assumption 2 above, the matrix ${\mathbf H}^H{\mathbf H}$ is the
Wishart distribution and as result, ${\mathbf F}^H{\mathbf
H}^H{\mathbf H}{\mathbf F}$ is also subject to the Wishart distribution~\cite{anderson1971, muirhead1982}.
Therefore, the inverse~(\ref{inverse}) almost exists if the number of receiving antennas is not
less than that of transmitting antennas~\cite{anderson1971, muirhead1982}.
In practice, the complexity of ZF receiver can also be very high when $N_t$ is large~\cite{Fatema18SYSI, Debbah13JSAC}. However, ZF receiver, which needs much less antennas than the matched filter (MF) based approach for achieving the same error performance in moderate and high SNR regime, is also popular for massive MIMO in order to strike a balance between performance and complexity~\cite{Debbah13JSAC, Edfors17access}. To reduce the complexity of ZF receiver, some low-complexity approximation methods can be employed~\cite{DickISCAS2013, Jing18TCOM}. 	

%
The ZF detection is captured by the following two steps:
\begin{enumerate}
 \item  Perform ZF equalization. Multiplying both sides of equation~(\ref{model}) by
the pseudo-inverse $({\mathbf H}{\mathbf F})^{+}$, we get
\begin{align}
{\mathbf r}'={\mathbf s}+{\boldsymbol\xi}',
\end{align}
where ${\mathbf r}'=({\mathbf H}{\mathbf F})^{+}{\mathbf r}$ and
${\boldsymbol\xi}'=({\mathbf H}{\mathbf F})^{+}{\boldsymbol\xi}$.
Under Assumption 3, ${\boldsymbol\xi}'$ is the
circularly-symmetric complex Gaussian noise with covariance
$\sigma^2 ({\mathbf F}^H{\mathbf H}^H{\mathbf H}{\mathbf
F})^{-1}$;
 \item Perform a hard decision to obtain an estimate
$\hat{s}_k$ of $s_k$, i.e.,
\begin{align*}
\hat{s}_k={\rm arg}\;\min\limits_{s_k\in {\mathcal
S}}|r'_k-s_k|^2.
\end{align*}
\end{enumerate}

Since ZF equalizer is a memoryless detection, i.e., the decision on
the current symbol does not affect the decision on the next
symbol, the average SEP over one vector signal
${\mathbf s}$ is the arithmetic mean of all SEPs.


\section{Optimal Precoder to Minimize the Average Symbol Error Probability \label{sec4}}

Our primary purpose of this section is to first give an explicit convex region for which the optimally precoding matrix ${\mathbf F}$ that minimizes the average SEP of the ZF detector can be obtained and then to uncover some nice structures for the optimally precoded system, which naturally leads us to taking full advantage of the Szeg\"o's theorem for the systematic study of the asymptotic behaviour on the resulting error performance of massive MIMO systems.


\subsection{SEP Expressions for M-ary Signals and Explicit Convex Regions of Objective Functions}
We first study the average SEP for given preocoding matrix $\mathbf{F}$, where the average is taken over all the fading
coefficients. The proposed approach generalizes the optimal transmitter design of the ZF equalizer in~\cite{ding03} not only from the deterministic model to the statistical MIMO channel model, but also from the specific BPSK signal constellation to the general $M$-ary signal constellations including PAM, PSK and QAM signalling.
\begin{lemma}\emph{SEP expression:}\label{lemma:SEP}
The average SEP for $M$-ary PAM, PSK and QAM signals with precoding matrix $\mathbf{F}$ are given by
\begin{align}\label{eqn:avrsep}
\text{P}({\mathbf F})
=\frac{1}{N_t}\sum_{k=1}^{N_t}G\left([\left({\mathbf F}^H{\mathbf \Sigma}{\mathbf F}\right)^{-1}]_{kk}\right),
\end{align}
where $G(x)$ is convex for $0<x<T$, and we have:
\begin{enumerate}
	\item For PAM signals, $G(x)=G_{\text{PAM}}(x)= \frac{2(M-1)}{M \pi} \int_0^{\pi/2}\left(1+\frac{3\eta   }{(M^2-1)x\sin^2\theta}\right)^{-(N_r-N_t+1)}d\theta$, $T=T_{\rm PAM}=\frac{ 3\eta  (N_r-N_t) }{2(M^2 -1) }$.
	\item For PSK signals, $G(x)=G_{\text{PAM}}(x)= \frac{2(M-1)}{M \pi} \int_0^{\pi/2}\left(1+\frac{3\eta   }{(M^2-1)x\sin^2\theta}\right)^{-(N_r-N_t+1)}d\theta$, $T=T_{\rm PSK}=\frac{\eta(N_r-N_t) \sin^2{(\pi/M)}}{2}$.
	\item For QAM signals, $G(x)=G_{\text{QAM}}(x)=\frac{4(\sqrt{M}-1)} {\sqrt{M} \pi}\int_{\pi/4}^{\pi/2}\left(1+\frac{3\eta }{2(M-1) x \sin^2\theta}\right)^{-(N_r-N_t+1)}d\theta
	 + \frac{4(\sqrt{M}-1)} {M\pi}\int_{0}^{\pi/4} \left(1+\frac{3 \eta }{2(M-1) x \sin^2\theta}\right)^{-(N_r-N_t+1)}d\theta$, $T=T_{\rm QAM}=\frac{3\eta (N_r-N_t) }{4(M-1)}$.
\end{enumerate}
\end{lemma}

The proof is provided in Appendix\ref{appendix:lemmaSEP}.

To attain a unified expression, we now drop the subscripts of both $G(x)$ and $T$. Hence, $G(x)$ is convex if $0<x\le
T$. Correspondingly, the noise power associated with ${\mathbf F}$  (e.g., the variable of the objective function of Eq.~\eqref{eqn:avrsep}) satisfies the following condition,
\begin{eqnarray}\label{snr}
0<[({\mathbf F^H}{\mathbf\Sigma}{\mathbf F})^{-1}]_{kk}&\le&T\quad\textrm{for }k=1, 2, \ldots,N_t.
\end{eqnarray}

%
To develop an explicit constraint from~\eqref{snr}, we have the following proposition:
\begin{proposition}\label{prop:convexity}
	If $\mathbf F$ is restricted to be in set $\big\{{\mathbf F}: \zeta_1( {\mathbf F} ^H {\mathbf F})  \ge \frac{1}{\lambda_1 T} \big\}$ where $\zeta_1( {\mathbf F} ^H {\mathbf F})$ is the minimum eigenvalue of ${\mathbf F} ^H {\mathbf F}$, then the constraint in~\eqref{snr} is satisfied.
	
\end{proposition}

The proof is given in Appendix\ref{appendix:convexity}. Proposition~\ref{prop:convexity} requires that the minimal average transmitting power of the subchannels, $\zeta_1( {\mathbf F} ^H {\mathbf F})$, must be larger than certain predefined threshold that is related to the modulation signals, system SNR $\eta$ and channel statistics.  It is worth pointing out that, our result here identifies the conditions for the optimality of the precoder to a much general and simplified form compared with that of~\cite{ding03}.
Since $T$ is proportional to system SNR $\eta$ and $N_r-N_t$, constraint in Proposition~\ref{prop:convexity} is easy to satisfy in slightly high SNR regime, especially for the large MIMO system considered in this paper.

\subsection{SEP-optimal Precoders}
\begin{theorem}\label{th:opt-zf}
Let the eigenvalue decomposition of ${\mathbf\Sigma}$ be
${\mathbf\Sigma}={\mathbf W}{\mathbf\Lambda}{\mathbf W}^H$, where
${\mathbf W}$ is a unitary matrix, and ${\mathbf\Lambda}={\rm
diag}(\lambda_1, \lambda_2, \cdots, \lambda_{N_t})$ with
$\lambda_1\le\lambda_2\le\cdots\lambda_{N_t}$. If $\mathbf F$ is restricted to be in set $\big\{{\mathbf F}: \zeta_1( {\mathbf F} ^H {\mathbf F})  \ge \frac{1}{\lambda_1 T} \big\}$,
where $T$ is the threshold in~\eqref{snr} and $\zeta_1( {\mathbf F} ^H {\mathbf F})  $ is the minimum eigenvalues of ${\mathbf F} ^H {\mathbf F} $. Then, the optimal precoder minimizing the average SEP is given by
\begin{eqnarray}\label{opsolu}
\widetilde{\mathbf F}=\frac{1}{\sqrt{\rm tr({\mathbf \Lambda}^{-1/2})}}{\mathbf W}{\mathbf \Lambda}^{-1/4}\widetilde{\mathbf V}_2,
\end{eqnarray}
where $\widetilde{\mathbf V}_2$ is the $N_t\times N_t$ normalized DFT matrix, and  the resulting minimum average SEP is determined by
\begin{eqnarray}\label{avrsep}
{\text {P}}_{\text{min}}(\widetilde{\mathbf F})=G\bigg ({\Big (\sum_{k=1}^{N_t}
\lambda_k^{-1/2} \Big )^2}/{N_t} \bigg ).
\end{eqnarray}

\end{theorem}
The proof is provided in Appendix\ref{append:theoremoptzf}.
Here, we make the following two comments on Theorem~\ref{th:opt-zf}.
\begin{enumerate}
\item The optimal precoder design problems with ZF detection were also considered in~\cite{ding03, kiessling04}. However, our Theorem~\ref{th:opt-zf} provides an explicitly sufficient condition that guarantees the optimality of the proposed precoder. Note that although the convexity constraints given in Proposition~\ref{prop:convexity} are rather mild, the design of precoding matrix $\mathbf{F}$ given in Theorem~\ref{th:opt-zf} can be suboptimal when these constraints are violated  as the objective function may not be convex.
\item Here, it is highly worth pointing out that the resulting SEP for the optimal precoder exposes a very interesting structure which motivates us to systematically study the asymptotic SEP performance in massive MIMO systems.
\end{enumerate}

\subsection{Extension to Multiuser Cases}
The main objective of this subsection is to show how to extend our design to multiuser massive MIMO systems. We consider the case where $K$ users each with $N_t$ transmitting antennas transmit to the base station with $N_r$ receiving antennas such that $N_r- K N_t\gg 1$.
The input and output relationship can be modeled by
\begin{align}
\bar{\mathbf{r}}=\bar{\mathbf{H}}\bar{\mathbf{F}}\bar{\mathbf{s}}\end{align}
in which $\bar{\mathbf{H}}=[\mathbf{H}_1, \mathbf{H}_2,\ldots, \mathbf{H}_K]$, $\bar{\mathbf{F}}={\rm diag}\{\mathbf{F}_1, \mathbf{F}_2\ldots, \mathbf{F}_K\}$, and $\bar{\mathbf{s}}=[\mathbf{s}_1^T, \mathbf{s}_2^T, \ldots, \mathbf{s}_K^T]^T$.
For the multiuser case, we assume that the channel are correlated for themselves but uncorrelated between each other since they are geographically separated, with covariance matrix ${\mathbf I}\otimes{\boldsymbol\Sigma}_k$ for the channel matrix $\mathbf{H}_k$. Since the derivations of the SEP expressions for PAM, PSK and QAM are similar, in the following discussion, we simply take the PAM constellation as an example.
We also notice that ${\mathbb E} [\bar{\mathbf F}^H \bar{\mathbf H}^H\bar{\mathbf H}\bar{\mathbf F}]={\mathbb E}[\bar{\mathbf F}^H \bar{\mathbf \Sigma}\bar{\mathbf F}]$, where $\bar{\mathbf \Sigma}={\rm diag}\big\{\mathbf{\Sigma}_1, \mathbf{\Sigma}_2,\ldots, \mathbf{\Sigma}_K\big\}$, and also
\begin{align}\label{eqn:corrinverse}
\!\left(\bar{\mathbf F}^H \bar{\mathbf \Sigma}\bar{\mathbf F}\right)^{-1}\!=\!{\rm diag}\big\{ [\mathbf{F}_1^H \mathbf{\Sigma}_1\mathbf{F}_1]^{-1},
\ldots, [\mathbf{F}_K^H \mathbf{\Sigma}_K\mathbf{F}_K]^{-1} \big\}.
\end{align}

For the considered multiuser system, the average SEP over all the users with precoding matrix $\bar{\mathbf{F}}$ is given by:
\begin{align}
&\text{P}_{\text{PAM}}(\bar{\mathbf F})
=\frac{2(M-1)}{M K N_t \pi}\sum_{k=1}^{K N_t }\int_0^{\frac{\pi}{2}}...\nonumber\\
&\qquad \mathbb{E}_{\bar{\mathbf H}}\text{exp}\left(-\frac{3\eta } {(M^2-1)[\left(\bar{\mathbf F}^H \bar{\mathbf H}^H{\mathbf H}{\mathbf F}\right)^{-1}]_{kk}\sin^2\theta}\right)\,d\theta \nonumber \\
&\stackrel{(a)}{=}\frac{2(M-1)}{M K N_t \pi}\sum_{k=1}^{K N_t}\int_0^{\frac{\pi}{2}} ....\nonumber\\
&\qquad \mathbb{E}_{\bar{\gamma}}\text{exp}\left(-\frac{3 \eta  \bar{\gamma} } {(M^2-1)[\left(\bar{\mathbf F}^H \bar{\mathbf \Sigma}\bar{\mathbf F}\right)^{-1}]_{kk}\sin^2\theta}\right)\,d\theta \nonumber \\
&=\frac{2(M-1)}{M K N_t \pi \Gamma(N_r-K N_t +1)}\sum_{k=1}^{K N_t} \int_0^{\frac{\pi}{2}} \int_0^\infty...\nonumber\\
&\text{exp}\!\left[\!-\!{\left(1\!+\!\frac{3\eta
	} {(M^2-1)[\left(\bar{\mathbf F}^H\bar{\mathbf \Sigma}\bar{\mathbf F}\right)^{-1}]_{k}\sin^2\theta} \right)} \bar{\gamma} \right] \!\bar{\gamma}^{N_r-K N_t} d\gamma d\theta \nonumber\\
\end{align}
where in $(a)$, likewise in~\eqref{eqn:chipdf}, we have used the fact that $\bar{\gamma}= \frac{[\left(\bar{\mathbf F}^H \bar{\mathbf \Sigma}\bar{\mathbf F}\right)^{-1}]_{kk} }{[\left(\bar{\mathbf F}^H \bar{\mathbf H}^H \bar{\mathbf H}\bar{\mathbf F}\right)^{-1}]_{kk}}$ is subject to ${\mathcal X}^2_{2(N_r-K N_t+1)}$. Now, we have
\begin{align}
&\text{P}_{\text{PAM}}(\bar{\mathbf F})\text{P}_{\text{PAM}}(\bar{\mathbf F})=\frac{2(M-1)}{M K N_t \pi}\sum_{k=1}^{K N_t}\int_0^{\frac{\pi}{2}}...\nonumber\\
&\left(1+\frac{3 \eta  }{(M^2-1)[\left(\bar{\mathbf F}^H \bar{\mathbf \Sigma}\bar{\mathbf
		F}\right)^{-1}]_{kk}\sin^2\theta}\right)^{-(N_r-K N_t+1)}d\theta \nonumber\\
&\stackrel{(b)}{=}\frac{1}{K N_t} \sum_{\ell=1}^K\sum_{k=1}^{N_t} {\bar G}_{\text{PAM}}\left([\left(\bar{\mathbf
	F}_\ell^H \bar{\mathbf\Sigma}_\ell\bar{\mathbf F}_\ell\right)^{-1}]_{kk}\right),\label{eqn:pamavrsepmultiuser}
\end{align}
we have used~\eqref{eqn:corrinverse} in $(b)$ and ${\bar G}_{\text{PAM}}(x)= \frac{2(M-1)}{M \pi} \int_0^{\frac{\pi}{2}}\left(1+\frac{3\eta   }{(M^2-1)x\sin^2\theta}\right)^{-(N_r-K N_t+1)}d\theta$. We can see from~\eqref{eqn:pamavrsepmultiuser} that the SEP over $K$ user is simply the arithmetic mean of the SEPs of all the users. As each user is subject to their individual power constraint, the precoder design can be decomposed into $K$ design problems, which can all be optimally solved by Theorem~\ref{th:opt-zf}.

\section{Asymptotic SEP Analysis for Optimally Precoded massive MIMO Systems}
In this section, our main purpose is to investigate the asymptotic behavior of SEP for the optimally precoded correlated massive MIMO systems equipped with the ZF receiver. The array size of both the transmitter and the receiver is increased while maintaining a constant ratio between them.

\subsection{Array Correlation Model with Toeplitz Covariance Matrix}
In general, MIMO techniques can yield linear increasing in the data rate against the minimum number of the transmitter and the receiver antennas in rich scattering environment, particularly when the array elements are uncorrelated~\cite{telatar95}. However, in a practical radio propagation process, correlation is almost inevitable, especially for a massive MIMO architecture.  In this paper, we assume that the transmitter array is arbitrarily correlated and that the correlation between each element of the receiver array is negligible.  This case can be considered as a MIMO system in the uplink where the transmitter is a mobile terminal with correlated array and the receiver is a base station, where the distance between adjacent antenna elements can be made as large as desired to eliminate correlation. To facilitate our analysis, we also assume that the correlation matrix is a Hermitian Toeplitz matrix~\cite{Loyka01}. This is a simplified model of measurement in practical environment, but it can capture the main phenomenon of spatial correlation between antennas~(see e.g., \cite{Foschini00mar} for other models). This model enables us to completely take advantage of the structure provided by the optimal system as well as of the Szeg\"o's theorem on large Hermitian Toeplitz matrices so that we can attain a simple closed-form solution in terms of the correlation coefficients, from which some important  insightful information on the effect of correlation can be extracted.

\subsection{Asymptotic Behaviour of Large Toeplitz Matrices}

To fully make use of the optimal structure provided by~\eqref{avrsep} for our analysis on the asymptotic behavior of the statistical average SEP, let us review an important property on a sequence of large Hermitian Toeplitz matrices $\{{\mathbf T}_K\}_{K=1}^{\infty}$.
Without loss of generality, we let
\begin{align}
{\mathbf T}_K = \begin{bmatrix}
t(0) & t(-1)&\cdots& t(-(K-1))\\
t(1) & t(0)&\cdots& t(-(K-2))\\
\vdots&\vdots&\ddots&\vdots  \\
t(K -1)& t(K-2) &\cdots&t(0)
\end{bmatrix}.
\end{align}
where $t(k)=t^*(-k)$ and $t(k)$ are assumed to be absolutely square-summable, i.e., $\sum_{k=0}^{\infty} |t(k)|^2 <\infty$. Thus, the following pair of discrete-time Fourier transforms exists,
\begin{align*}
s_{\mathbf T} (\omega) &= \sum_{k=-\infty}^\infty t(k) e^{-jk\omega},\label{eqn:spfconvg}\quad
t(k) = \frac{1}{2\pi}\int_0^{2\pi} s_{\mathbf T}(\omega) e^{jk\omega} d\omega.
\end{align*}
It is worth noting that the function $s_{\mathbf T}(w) $ is real, since $\mathbf T$ is Hermitian and  $s_T(w)$ is also known as the power spectral density (PSD) function. The above relationship is also known as the Wiener-Khinchin theorem of discrete-time process.

\begin{lemma}[Szeg\"o's theorem]\label{thm:szeg}~\cite{Gray06}
Let $\{{\mathbf T}_K \}_{K=1}^\infty$be a sequence of Hermitian Toeplitz matrices with $K$ eigenvalues of ${\mathbf T}_K$ given by $\mu_{K,1}\le \mu_{K, 2} \le \cdots \le \mu_{K, K}$, and $\sum_{k=0}^{\infty}|t(k)|^2$ being convergent. Then for any function ${\mathtt F}(x)$  that is continuous on $[L_{s_{\mathbf T}},  U_{s_{\mathbf T}}]$, where $L_{s_{\mathbf T}} = \text{ess~inf}~s_{\mathbf T}(\omega)$ is the essential infimum~\cite{Royden10} of $s_{\mathbf T}(\omega)$ and defined to be the largest value of $c$ for which $s_{\mathbf T}(\omega) \ge c$ except on a set of measure 0, and $U_{s_{\mathbf T}} = \text{ess~sup}~s_{\mathbf T}(\omega)$ is the smallest number $d$ for which $s_{\mathbf T}(\omega) \le d$ except for a set of measure 0, we have
\begin{align}
\lim_{K\to \infty} \frac{1}{K} \sum_{\ell=1}^{K} {\mathtt F}(\mu_{K, \ell}) =\frac{1}{2\pi} \int_{0}^{2\pi} {\mathtt F}(s_{\mathbf T}(\omega)) d\omega.
\end{align}
~
\end{lemma}
The above Lemma~\ref{thm:szeg} plays a vital role in the asymptotic error performance analysis of the considered massive MIMO system.

\subsection{Asymptotic SEP Analysis for Massive MIMO with Toeplitz Covariance Matrix}
We are now ready to present the asymptotic SEP analysis for the precoded massive MIMO system.  From now on, we assume that the ratio of the number of the receiver antennas to that of the transmitter antennas is fixed, i.e., $N_r/N_t=\beta > 1$ is constant. The average SEP-optimal coding design fits both conventional and massive MIMO. However, an important question is how the SEP behaves when the number of transmitting antennas goes to infinity while keeping $\beta=\frac{N_r}{N_t} $ and the transmitting power fixed. By strategically resorting to the Sezg\"o's theorem, we manage to show that the SEP quickly converges to a fixed value when the system SNR $\eta$ is fixed. The main result of this paper can be formally stated as the following theorem.

\begin{theorem}\label{szego:sep}
Let us consider massive MIMO systems using the optimal precoder in~\eqref{opsolu}, the ZF detector and the $M$-ary PAM, PSK or QAM constellations. If the entries of the channel covariance matrix ${\mathbf \Sigma}$ are absolutely square-summable and the resulting $L_{s_{\Sigma}}>0$, then, $\lim_{N_t\to\infty} {\rm P}_{N_t}(\widetilde{\mathbf F})=\bar{{\rm P}}_{\rm opt}$ exists and
\begin{itemize}
\item
$\bar{{\rm P}}_{\rm opt, PAM} = \frac{2(M-1)}{M} Q\left( \sqrt{  \frac{ 6\eta (\beta-1) }{(M^2-1)\Lambda^2 }}\right)$;
\item
$\bar{{\rm P}}_{\rm opt, PSK} =\frac{1}{\pi} \int_0^{ (M-1)\pi/M } \exp \left( - \frac{\eta (\beta-1) \sin^2(\pi/M)}{\Lambda^2\sin^2\theta} \right) d\theta$;
\item
$\bar{{\rm P}}_{\rm opt, QAM} =\frac{4(\sqrt{M}-1)}{\sqrt{M}}Q\left(\sqrt{\frac{3\eta(\beta-1) }{(M-1)\Lambda^2}} \right) -\frac{4(\sqrt{M}-1)^2}{M}  Q^2\left( \sqrt{ \frac{3\eta(\beta-1) }{(M-1) \Lambda^2}} \right)$.
\end{itemize}
where $\Lambda$ is defined by $\Lambda= \frac{1}{2\pi} \int_{0}^{2\pi} \frac{d\omega}{\sqrt{s_{\Sigma}(\omega)}} $  with $s_{\Sigma}(\omega)=\sum_{k=-\infty}^{\infty} \sigma(k) e^{-jk\omega}$, where $\sigma(m-n) =[{\bf \Sigma }]_{mn}$ denotes the $(m,n)$-th entry of $\mathbf{\Sigma}$.~
\end{theorem}

The proof can be found in Appendix\ref{append:szegosep} and we would like to make the following two comments:
\begin{enumerate}
\item From Theorem~\ref{th:opt-zf} we can see that the diversity gain for the optimally precoded MIMO system for a fixed $N_t$ with the ZF receiver is $N_r-N_t+1$. However, when $N_t$ tends to infinity, Theorem~\ref{szego:sep} reveals that the limiting SEP of the optimally precoded MIMO system equipped with the ZF detector decays exponentially.
\item Despite the fact that the assumption of Theorem~\ref{szego:sep} requires that the correlation matrix is Toeplitz so as to make use of the Szeg\"o's theorem, we can infer from the following proof that the assumption can be actually relaxed to any invertible correlation matrix ${\mathbf\Sigma}$ with the condition that $\lim _{N_t\to \infty} \frac{1}{N_t}\sum_{n=1}^{N_t}\lambda_k^{-1/2}$ exists, where $\lambda_1\le \lambda_2\le\cdots\le \lambda_{N_t}$ are the eigenvalues of ${\mathbf\Sigma}$.
\end{enumerate}

We now show that Theorem~\ref{szego:sep} can be used for the SEP evaluation of precoded massive MIMO with correlated antennas. In particular, when the channel covariance matrix ${\bf \Sigma }$ is the non-symmetric Kac-Murdock-Szeg\"o~(KMS) matrix that has been used widely in the literature ~\cite{KMS53, Turin62, Loyka01, Veeravalli01, Mestre03, Dow2003, Win08, Gesbert11, Zoltowski14}, the $(m,n)$-th entry of which is denoted by $\sigma(m-n)$, i.e.,
\begin{align}\label{kmsmatrix}
\sigma(m-n) =[{\bf \Sigma }]_{mn}  =\left\{ \begin{matrix}
\rho^{n-m}& m \le n,\\
[{\mathbf \Sigma}]^{*}_{nm}&m>n,
\end{matrix} \right.
\end{align}
where $ 0 < |\rho| < 1$  indicates the degree of correlation, we have the following corollary.
\begin{corollary}\label{corollary:kms}
Consider massive MIMO systems with the optimal precoder~\eqref{opsolu}, ZF detector and the $M$-ary PAM, PSK and QAM constellation. If $ 0 < |\rho| < 1$, then, $\lim_{N_t\to\infty} {\rm P}_{N_t}(\widetilde{\mathbf F})=\widetilde{\rm P}$ exists and
\begin{align*}
&\widetilde{\text{P}}_{{\rm PAM}}= \frac{2(M-1)}{M} Q\left( \sqrt{  \frac{ 3 \pi^2 \eta (1-|\rho|)(\beta-1) } {2(1+|\rho|)(M^2-1)  {\mathtt E}^2\left( \frac{2\sqrt{|\rho|} }{ 1+|\rho|}\right) }}\right);\\
&\widetilde{\text{P}}_{{\rm PSK}}\!=\frac{1}{\pi}\!\int_0^{ \frac{(M-1)\pi}{M}}\!\!\!\!\!\exp \left(\frac{-\pi^2\eta (1-|\rho| )(\beta-1) \sin^2(\frac{\pi}{M})}{4(1+|\rho|) {\mathtt E}^2\left( \frac{2\sqrt{|\rho|} }{ 1+|\rho|}\right) \sin^2\theta} \right) d\theta;\\
&\widetilde{\text{P}}_{{\rm QAM}}=
\frac{4(\sqrt{M}-1)}{\sqrt{M}} Q\Bigg(\sqrt{\frac{ 3\pi^2 \eta(1-|\rho|)(\beta-1) }{ 4(1+|\rho|)(M-1) {\mathtt E}^2\Big( \frac{2\sqrt{|\rho|} }{ 1+|\rho|}\Big) }} \Bigg)
\\
&-\frac{4(\sqrt{M}-1)^2}{M}  Q^2\Bigg(\sqrt{\frac{ 3\pi^2 \eta(1-|\rho|)(\beta-1) }{ 4(1+|\rho|)(M-1) {\mathtt E}^2\Big( \frac{2\sqrt{|\rho|} }{ 1+|\rho|}\Big) }} \Bigg),
\end{align*}
where ${\mathtt E}(k) =\int_0^{\pi/2}\sqrt{1-k^2\sin^2 \theta}\,d\theta$ denotes the complete elliptic integral of the second kind~\cite{Gradshteyn}.~
\end{corollary}
The proof is provided in Appendix\ref{append:kms}.

\subsection{Convex Region for KMS Matrices}
Let $\zeta_1({\mathbf \Sigma})\le \zeta_2({\mathbf \Sigma}) \le \ldots \le \zeta_{N_t}({\mathbf \Sigma}) $ be the eigenvalues of  the KMS matrix and then from~\cite{Yueh05eigenvaluesof}, we have $\zeta_k({\mathbf \Sigma}) = \frac{1-|\rho|^2}{1+ |\rho|^2+2 |\rho| \cos \theta_k},~\text{for } k=1,2,\ldots, N_t$,
where $\cos\theta_1 \ge \cos\theta_2 \ge \ldots \ge \cos\theta_{N_t}$, in which  $\theta_k$ is the solution to
$|\rho|^2 \Big(\sin(N_t+1)\theta_k + 2|\rho|\sin N_t\theta_k +|\rho|^2 \sin(N_t-1)\theta_k\Big) =0$.
Since $|\cos \theta_k| \le 1$ and $0<|\rho|<1$, then for the KMS matrix, $\zeta_1({\mathbf \Sigma}) \ge  \frac{1-|\rho|^2}{(1+|\rho|)^2} =\frac{1-|\rho|}{1+|\rho|} >0$.
Recall that the optimality condition for the precoder is
\begin{align}
\zeta_1({\mathbf F}^H {\mathbf F}) \ge \frac{1}{\zeta_1(\mathbf \Sigma) T}
\end{align}
where $\zeta_1({\mathbf F}^H {\mathbf F})$ is the minimum eigenvalue of ${\mathbf F}^H {\mathbf F}$. Finally, for KMS covariance matrix, the constraint~\eqref{newcvxconstraint} have the following simple sufficient form
\begin{align*}
\zeta_1 ({\mathbf F}^H {\mathbf F})\ge \frac{ 1+|\rho|}{T(1-|\rho|)}.
\end{align*}

\subsection{Limiting Performance of the Individual SNR}
To further appreciate the asymptotic SEP properties derived for the optimally precoded massive MIMO systems in the previous subsection, we are also motivated to study the asymptotic distribution of the SNR for each sub-channel when the array size is large.  Notice that at the output of the ZF receiver for each sub-channel, the average signal power is $\mathbb E[|s_k|^2]=1$, and the power of the equalized noise is $\sigma^2
\left[ (\widetilde{\mathbf F}^H{\mathbf H}^H{\mathbf H}\widetilde{\mathbf F})^{-1}\right]_{kk}$.
Therefore, the instantaneous SNR of each
sub-channel as a function of the random channel realization is
\begin{align*}
	\tau_k =\frac{1}{ \sigma^2 \left[  (\widetilde{\mathbf F}^H{\mathbf H}^H{\mathbf H}\widetilde{\mathbf F})^{-1}\right]_{kk}}\quad\text{for } k=1,2,\ldots, N_t,
\end{align*}
Now, we have the following remark on $\tau_k$ when $N_t$ goes to be unlimited.
\begin{remark}\label{remarklimitsnr} For the asymptotic behaviour on individual SNR for each subchannel, we have $\lim_{N_t \to \infty} \tau_k \xrightarrow{\text{ almost surely}}   \frac{\eta(\beta-1) }{\Lambda^2} {~\rm for~} k=1,2,\ldots,N_t$. 	
\end{remark}
\emph{Proof:} First, by Assumption 5, we have $\eta=1/\sigma^2$ and hence $\tau_k =\frac{\eta}{ \left[  (\widetilde{\mathbf F}^H{\mathbf H}^H{\mathbf H}\widetilde{\mathbf F})^{-1}\right]_{kk}}
	=\frac{N_t \eta \widetilde{\gamma}_k}{ \left(\sum_{m=1}^{N_t} \lambda_m^{-1/2}\right)^2},\quad\text{for } k=1,2,\ldots, N_t$,
where $\widetilde{\gamma}_k =\frac{[\left(\widetilde{\mathbf F}^H{\mathbf \Sigma}\widetilde{\mathbf F}\right)^{-1}]_{kk} }{[\left(\widetilde{\mathbf F}^H{\mathbf H}^H {\mathbf H}\widetilde{\mathbf F}\right)^{-1}]_{kk}}$ and $\widetilde{\mathbf F}$ is the optimal precoder given in~\eqref{opsolu}.
Therefore, the mean and variance of $\tau_k $ can be determined as follows:
\begin{align*}
	{\mathbb E}[\tau_k] &= \frac{N_t (N_r-N_t+1)\eta}{\left(\sum_{m=1}^{N_t} \lambda_m^{-1/2}\right)^2},~
	{\rm var} [\tau_k] = \frac{N_t^2 (N_r-N_t+1)\eta^2}{\left(\sum_{m=1}^{N_t} \lambda_m^{-1/2}\right)^4}. \nonumber
\end{align*}
When scaling up the array size, and with the help of~\eqref{eqn:szegolimit}, we have
\begin{align*}
	\lim_{N_t \to \infty}  {\mathbb E}[\tau_k]& =    \frac{\eta(\beta-1) }{\Lambda^2 },
	\lim_{N_t \to \infty}  {\rm var}[\tau_k]
	=\lim_{N_t \to \infty}  \frac{\beta-1}{ \sigma^4 \Lambda^4 N_t}
	= 0.
\end{align*}
Then, by the law of large numbers (LLN), we have
\begin{align*}
	\lim_{N_t \to \infty} \tau_k \xrightarrow{\text{ almost surely}}   \frac{\eta(\beta-1) }{\Lambda^2} {\rm~for~} k=1,2,\ldots,N_t.
\end{align*}

\begin{figure}[ht]
	\centering
	\flushleft
	\resizebox{8cm}{!}{\includegraphics{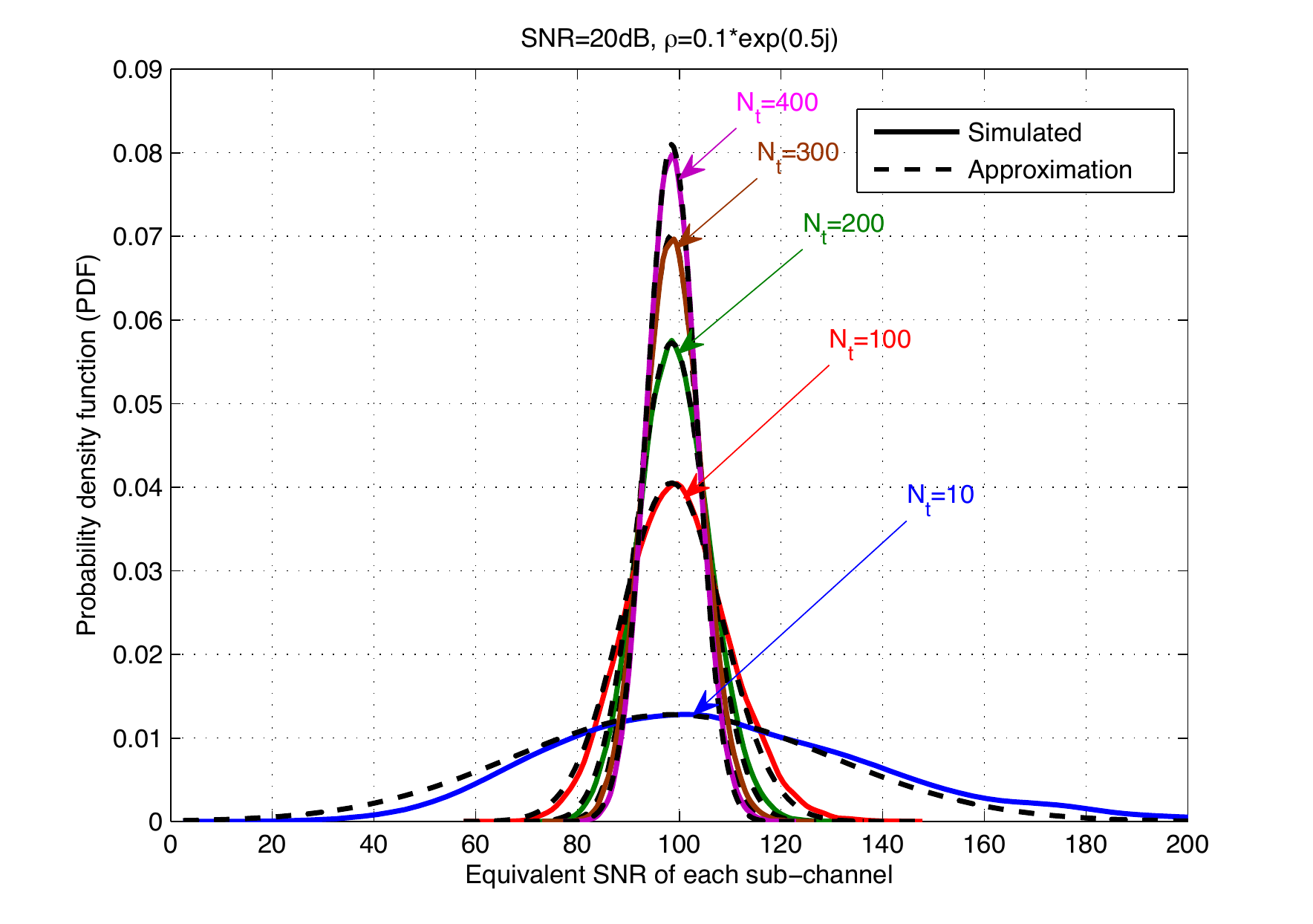}}
	\centering
	\caption{The distribution of the equivalent SNRs of each sub-channel with $\beta =2$, and $\rho=0.1*\exp(0.5j)$, SNR = 20dB.}
	\label{fig:asympspec1}
\end{figure}

\begin{figure}[ht]
	\centering
	\flushleft
	\resizebox{8cm}{!}{\includegraphics{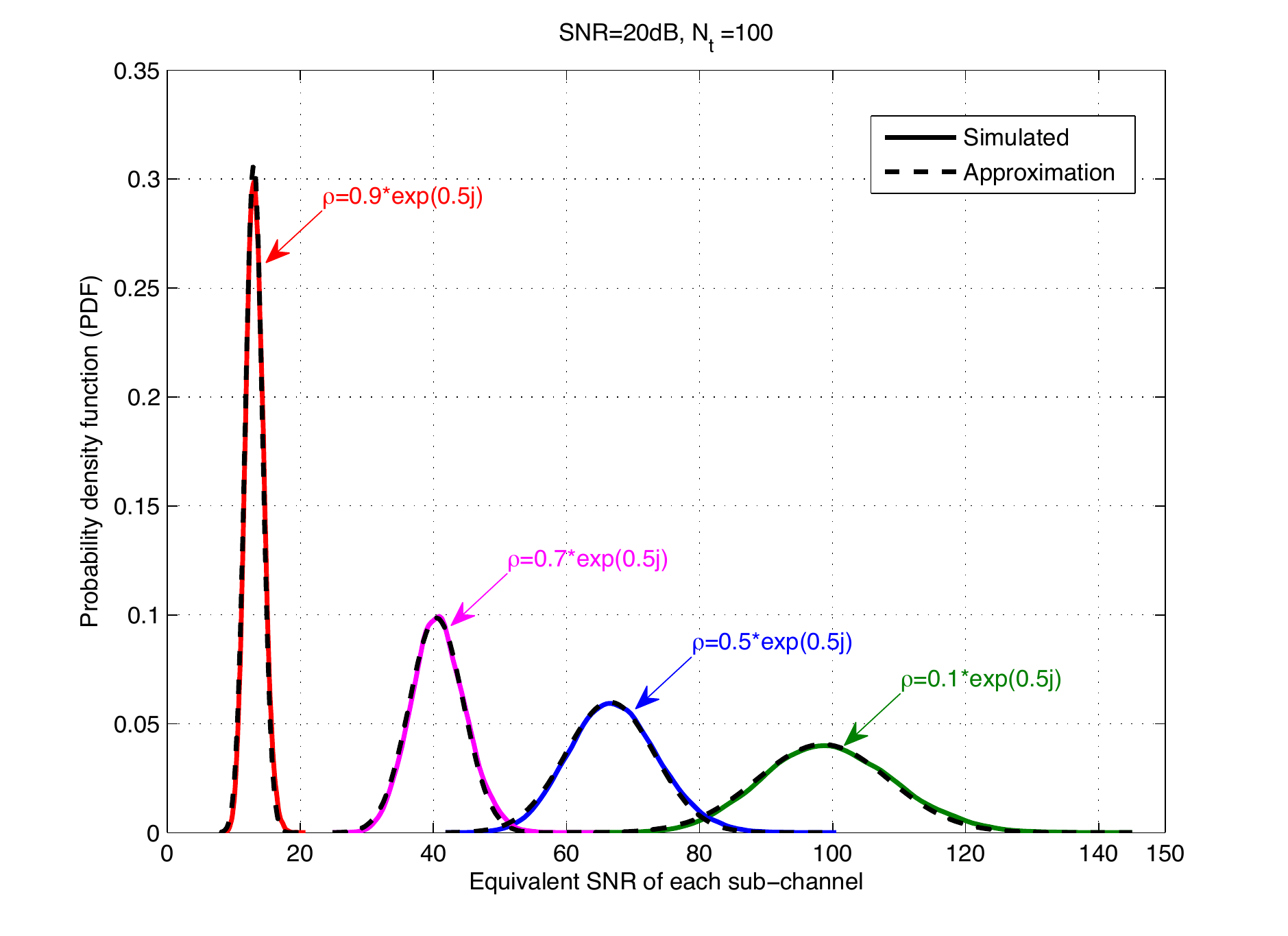}}
	\centering
	\caption{The distribution of the equivalent SNRs of each sub-channel with $\beta =2$, and $N_t =100$, SNR= 20dB.}
	\label{fig:asympspec2}
\end{figure}

Remark~\ref{remarklimitsnr} suggests that when the size of the antenna array goes to infinity, the instantaneous SNR of each sub-channel
becomes stable, i.e., it converges to a fixed value. This verifies the results in Theorem~\ref{szego:sep}.
Here, note that the exact convergence requires the array size goes to infinity and it does not necessarily work well when the array size is small. In what follows, we give an intuitive approximation to the distribution of the SNR for each receiver branch, which is very accurate when the array size is moderate large. From the convergence of Szeg$\ddot{\text{o}}$' Theorem in~\eqref{eqn:szegolimit}, we know that
\begin{align}
\lim_{N_t \to \infty} \tau_k = \lim_{N_t \to \infty} \frac{N_t \eta \widetilde{\gamma}_k}{ \left(\sum_{m=1}^{N_t} \lambda_m^{-1/2}\right)^2}\approx\frac{\eta\widetilde{\gamma}_k}{ N_t\Lambda^2}
\end{align}
for a large $N_t$. Now letting $\tilde \tau_k=\frac{\eta\widetilde{\gamma}_k}{ N_t\Lambda^2}$ and as the Szeg\"o's theorem converges very fast for the considered correlation matrix, $\tilde \tau_k \approx \tau_k$ when $N_t$ is reasonably large. Since $f(\widetilde{\gamma}_k) =\frac{1}{\Gamma(N_r-N_t +1)}e^{-\widetilde{\gamma}_k }\widetilde{\gamma}_k^{N_r-N_t } $, $\tilde \tau_k$ is subject to  the Gamma distribution with mean $ \frac{(N_r -N_t+1)\eta}{ N_t \Lambda^2}\approx \frac{\eta(\beta-1)}{ \Lambda^2}$ and variance $ \frac{  (N_r-N_t+1)\eta^2 }{ N_t^2 \Lambda^4} \approx\frac{ \eta^2(\beta-1)}{  N_t\Lambda^4 } $ when $N_t$ is large. Now, by the well-known central limit theorem, we have
\begin{align}\label{eqn:snrapprox}
\tilde \tau_k ~\dot \sim~ {\mathcal N} \left(\frac{\eta(\beta-1)}{\Lambda^2}, \frac{ \eta^2(\beta-1)}{  N_t\Lambda^4 } \right)
\end{align}
where $ \dot \sim$ means approximately with the same distribution when the array size is large.  Hence,
\begin{align*}
\lim_{N_t \to \infty} \tau_k
&=  \lim_{N_t \to \infty} \tilde \tau_k \sim ~{\mathcal N} \left(\frac{\eta(\beta-1)}{\Lambda^2},   \lim_{N_t \to \infty}  \frac{ \eta^2(\beta-1)}{  N_t\Lambda^4 }\right)\\
&\xrightarrow{\text{ almost surely}}   \frac{\eta(\beta-1) }{\Lambda^2}.
\end{align*}
It is worth pointing out that the approximation in~\eqref{eqn:snrapprox} is pretty accurate when the array size is relatively small, say, $N_t=10$, as can be seen in Figs.~\ref{fig:asympspec1} and~\ref{fig:asympspec2}.

\subsection{Uniform Power Allocation Strategy}
As a comparison, we are also interested with the system performance when no channel information is available at the transmitter. In this scenario, the transmitter cannot perform optimization on the input covariance matrix or carry out power allocation across transmitter antennas. Since in this case there would be no bias in terms of the mean or covariance of the channel matrix $\mathbf{H}$, the best precoding strategy would be to allocate equal power to each transmitter antenna and to make the covariance matrix omni-directional.
As a result, we consider asymptotic SEP for uniformly precoded massive MIMO channels, i.e., ${\hat{\mathbf F}}=\frac{1}{\sqrt{N_t}}{\mathbf I}$. We then have the following theorem:

\begin{theorem}\label{thm:uniformpallo}
Consider massive MIMO systems using the uniform precoder, ZF detector and the $M$-ary PAM, PSK and square QAM constellations. If the channel covariance matrix ${\mathbf \Sigma}$ is the KMS matrix in~\eqref{kmsmatrix}, then, $\lim_{N_t\to\infty} {\rm P}(\hat{\mathbf F})=\widehat{{\rm P}}_{\rm  U}$ exists and
\begin{align*}
&\widehat{{\rm P}}_{{\rm U, PAM}}= \frac{2(M-1)}{M } Q\left(\sqrt{\frac{6\eta(\beta-1)(1-|\rho|^2)}{(M^2-1)(1+ |\rho|^2) } } \right);\\
&\widehat{{\rm P}}_{{\rm U, PSK}}=\frac{1}{\pi} \int_0^{ \frac{(M-1)\pi}{M} } \!\!\!\!\exp \left( - \frac{ \eta(\beta-1) (1- |\rho|^2)\sin^2(\frac{\pi}{M})}{ (1+ |\rho|^2)\sin^2\theta} \right) d\theta;\\
&\widehat{{\rm P}}_{{\rm U, QAM}}=\frac{4(\sqrt{M}-1)}{\sqrt{M}} Q\left(\sqrt{\frac{3\eta(\beta-1) (1-|\rho|^2) }{(M-1)(1+|\rho|^2)  }} \right)\\
& -\frac{4(\sqrt{M}-1)^2}{M} Q^2\left( \sqrt{ \frac{3\eta(\beta-1)(1-|\rho|^2) }{(M-1) (1+|\rho|^2)   }} \right).
\end{align*}
~
\end{theorem}
The proof is provided in Appendix\ref{appendix:thmkms}.
 \begin{figure}[ht]
	\centering
	\flushleft
	\resizebox{8cm}{!}{\includegraphics{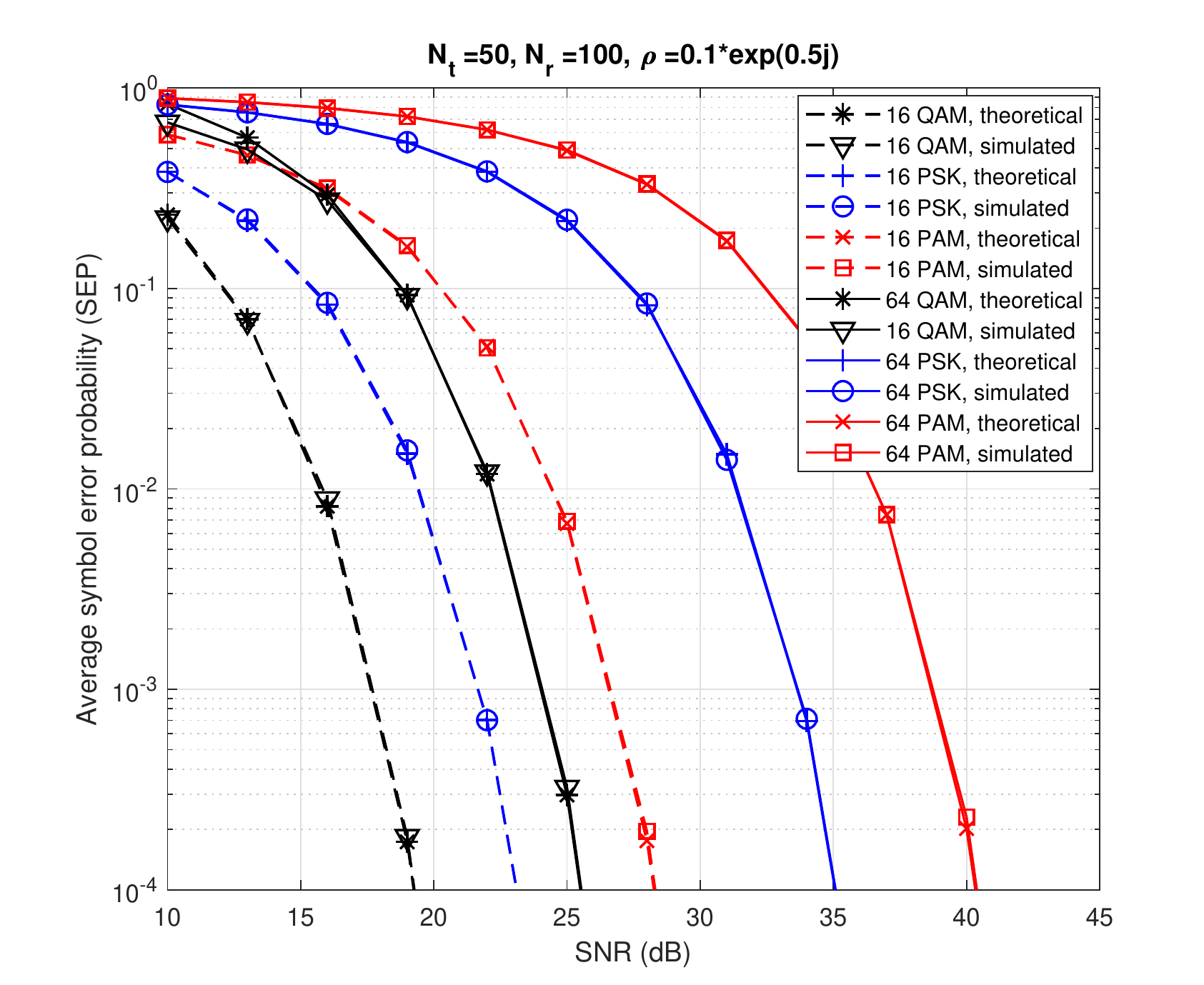}}
	\centering
	\caption{Average SEP performance  against SNR $\eta$, where the correlation coefficient $\rho=0.1*\exp(0.5j)$, $N_t=50$, $N_r=100$.
	}
	\label{fig:ser1}
\end{figure}
\begin{figure}[ht]
	\centering
	\flushleft
	\resizebox{8cm}{!}{\includegraphics{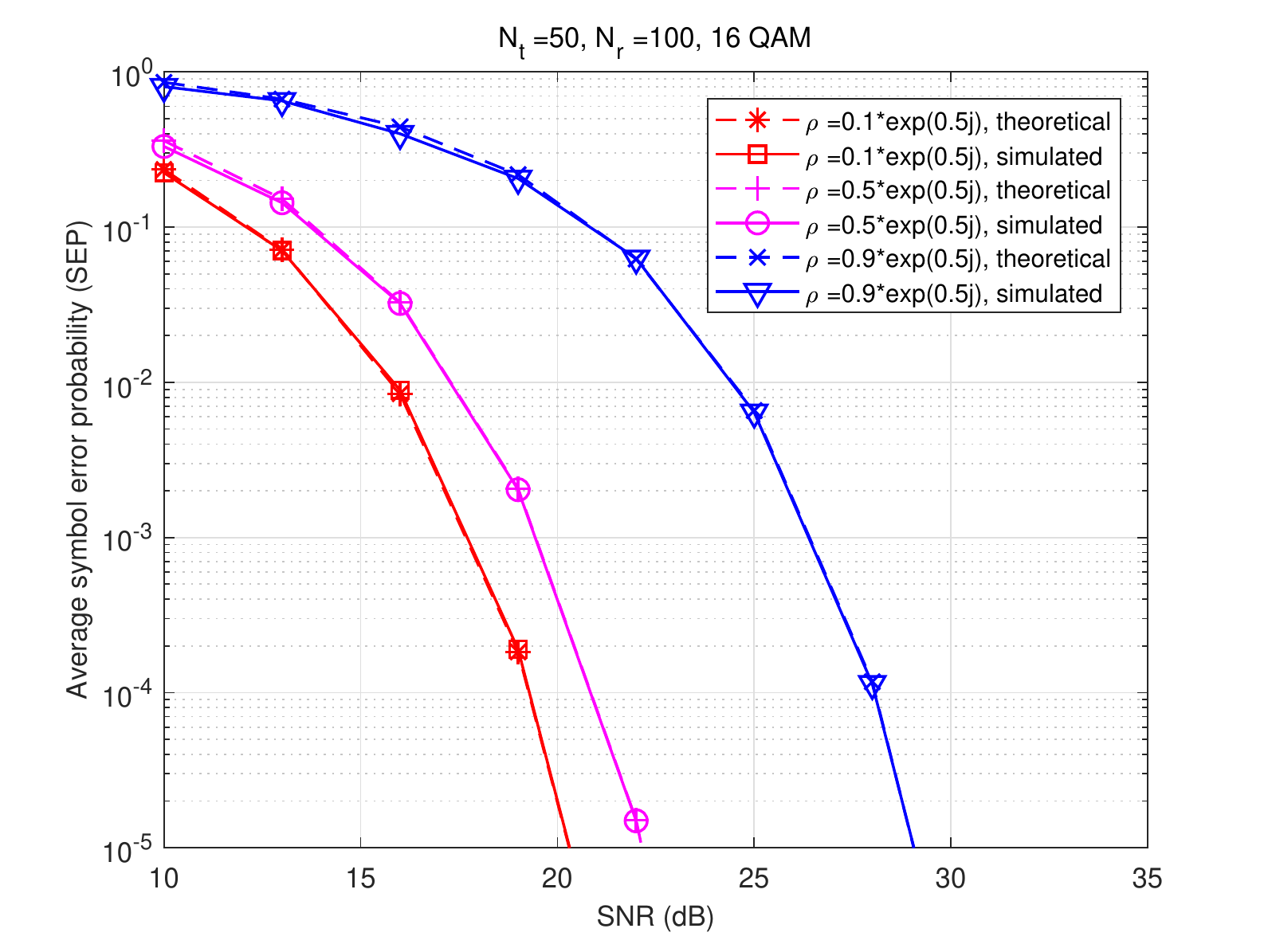}}
	\centering
	\caption{Average SEP performance against SNR $\eta$ with 16-QAM constellation.
	}
	\label{fig:ser2}
\end{figure}

\begin{figure}[ht]
	\centering
	\flushleft
	\resizebox{8cm}{!}{\includegraphics{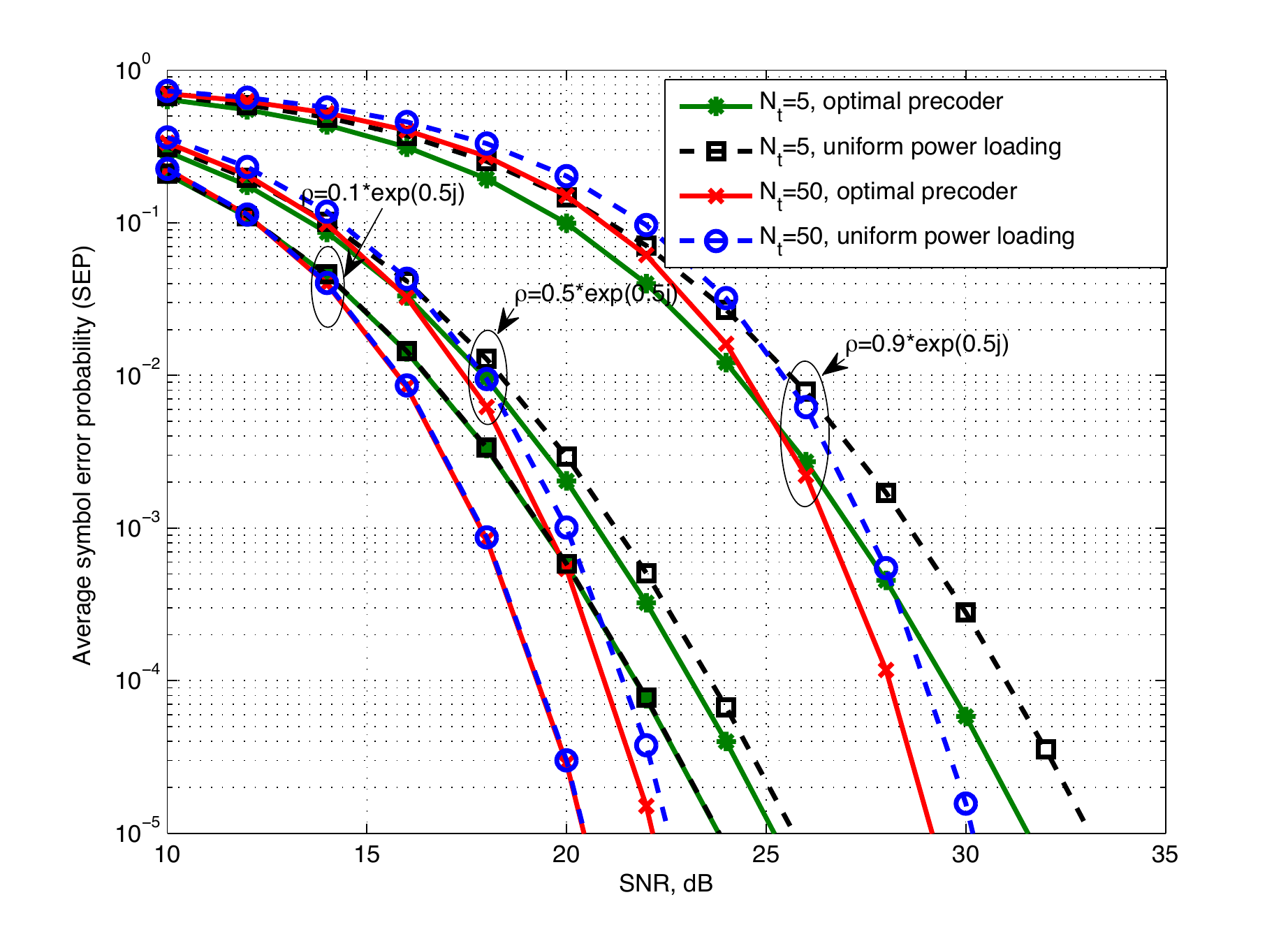}}
	\centering
	\caption{ Average SEP performance against SNR,  with 16-QAM, $\beta =2$ and different $\rho$. }
	\label{fig:unifpwrload}
\end{figure}

\begin{figure}[ht]
	\centering
	\flushleft
	\resizebox{8cm}{!}{\includegraphics{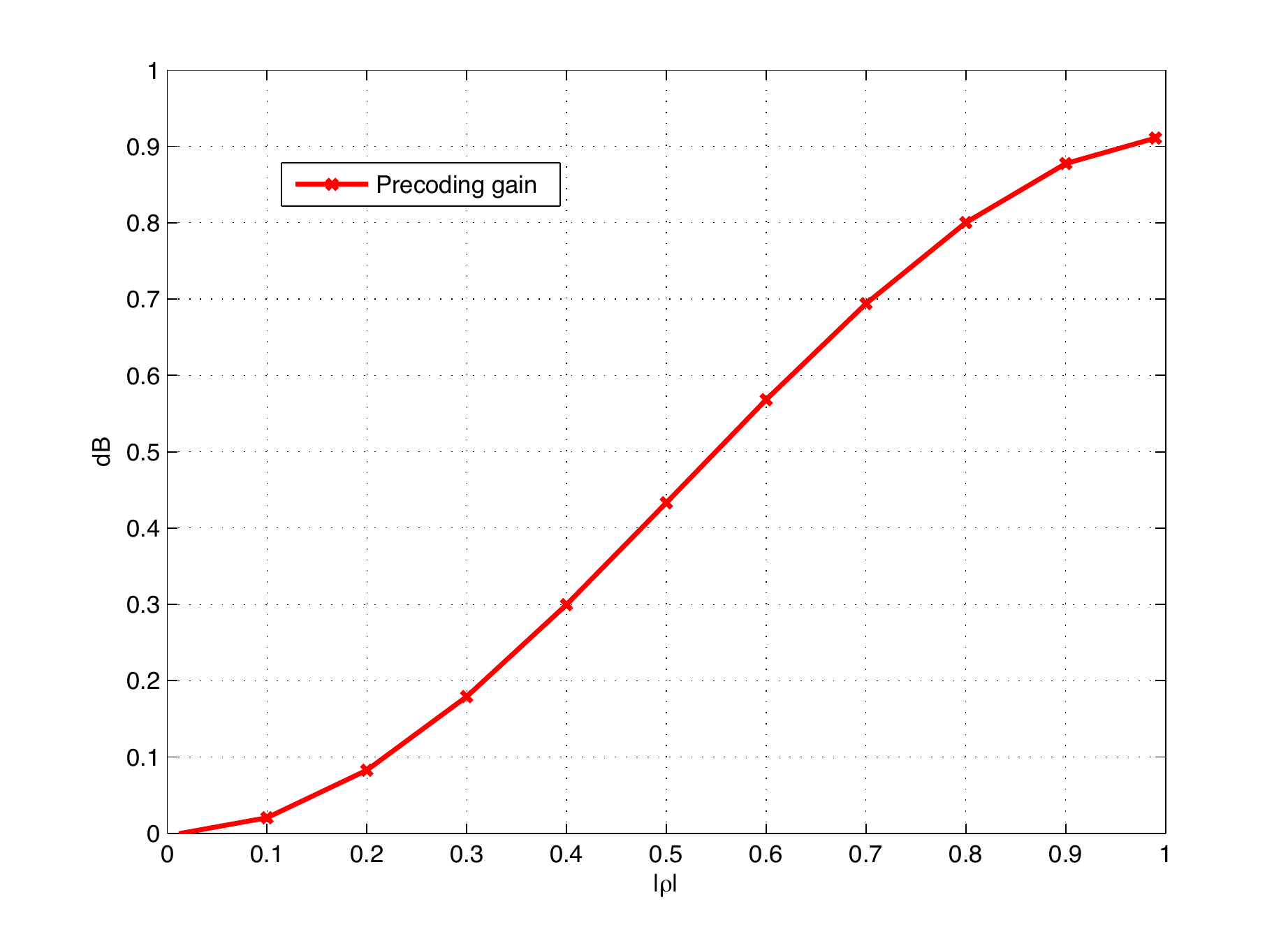}}
	\centering
	\caption{ The precoding gain compared with uniform power allocation versus $|\rho|$.}
	\label{fig:precodinggain}
\end{figure}

\begin{figure}[ht]
	\centering
	\flushleft
	\resizebox{8cm}{!}{\includegraphics{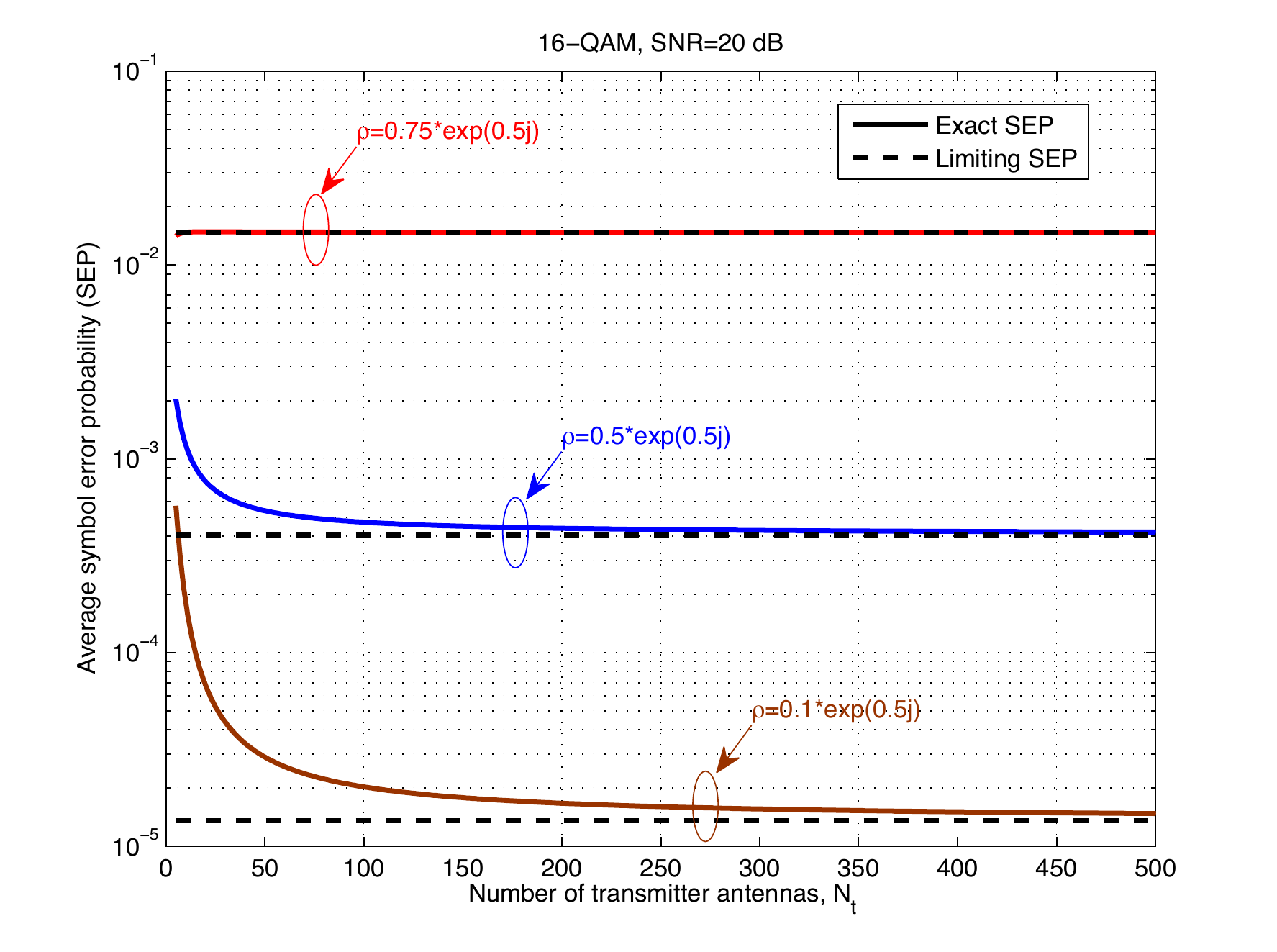}}
	\centering
	\caption{Average SEP performance against the number of transmitter antennas, with 16-QAM, $\beta =2$, and SNR = 20dB.}
	\label{fig:varrho}
\end{figure}

\begin{figure}[ht]
	\centering
	\flushleft
	\resizebox{8cm}{!}{\includegraphics{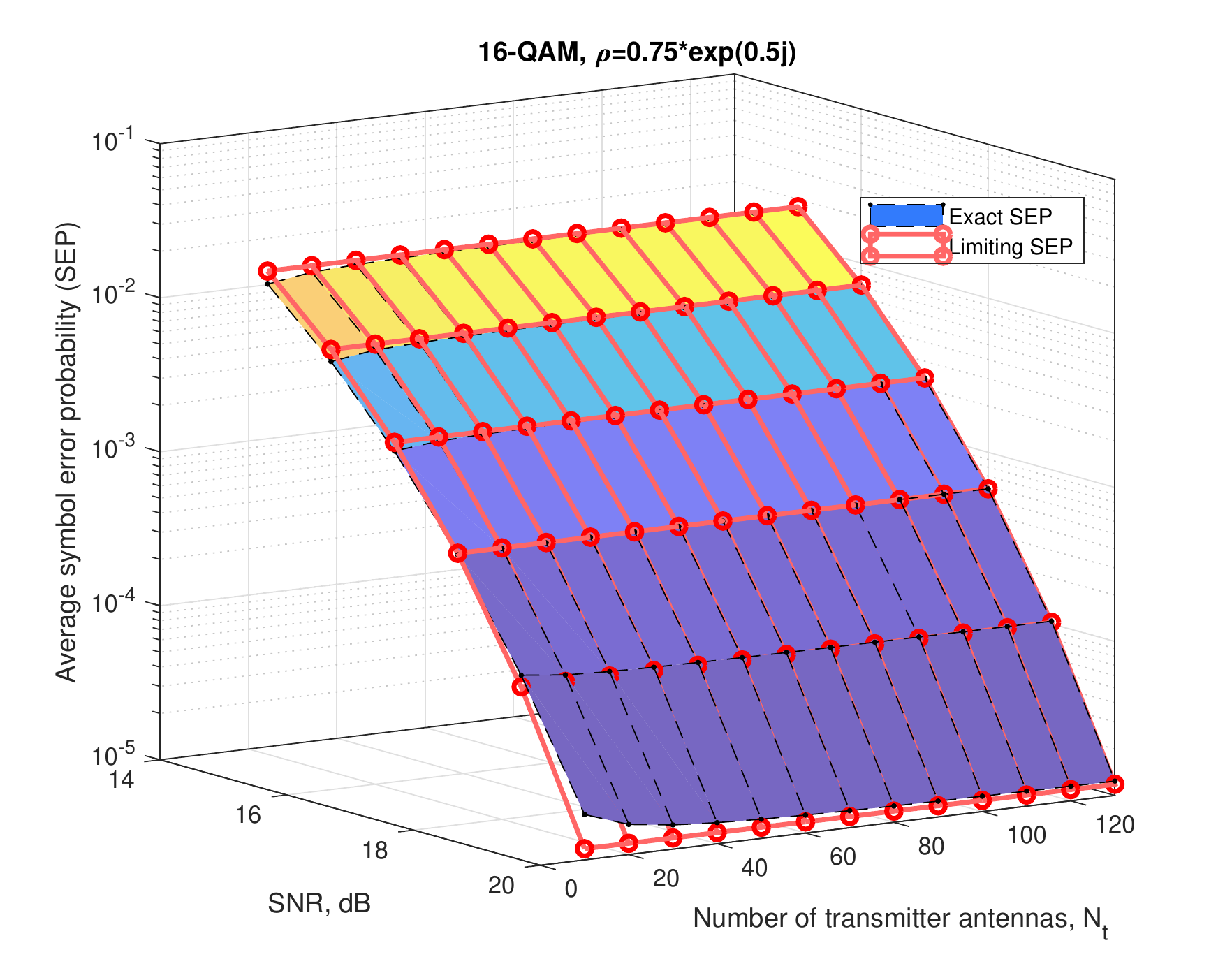}}
	\centering
	\caption{Average SEP performance against the number of transmitter antennas $N_t$ and SNR, with 16-QAM, $\beta =2$, and $\rho=0.75*\exp(0.5j)$.}
	\label{fig:varsnr}
\end{figure}
\begin{figure}[ht]
	\centering
	\flushleft
	\resizebox{8cm}{!}{\includegraphics{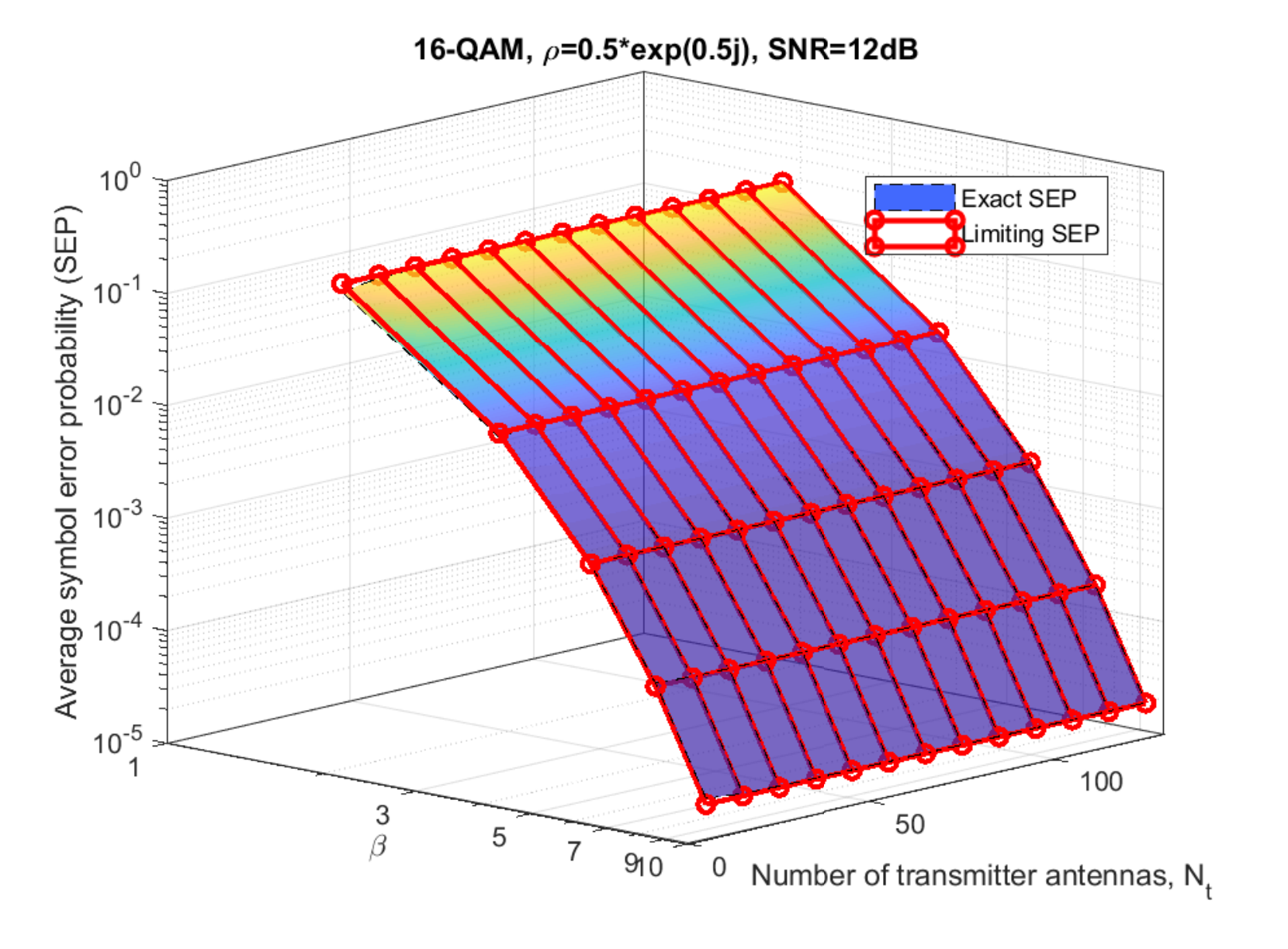}}
	\centering
	\caption{Average SEP performance against the number of transmitter antennas $N_t$ and $\beta$, with 16-QAM, $\rho=0.5*\exp(0.5j)$, and SNR= 12dB.}
	\label{fig:varbeta}
\end{figure}

\section{Numerical Simulations\label{sec7}}
In this section, we verify our theoretical results through computer simulations. In order to validate the theoretical SEP expression, Monte Carlo simulations are carried out. Let us first consider a uniform linear array with $N_t = 50$ transmitting antennas and $N_r=100$ receiving antennas, where the receiver knows the CSI perfectly and the transmitter knows only the correlation matrix $\bf \Sigma$.  In this simulation, the correlation matrix $\bf \Sigma$ is taken as the Kac-Murdock-Szeg\"o matrix.
The theoretical and the simulated SEP results for the optimal precoder are plotted in Fig.~\ref{fig:ser1} and Fig.~\ref{fig:ser2} with different constellations (PAM, PSK and square QAM) and channel correlation coefficients against the SNR $\eta$.
It can be observed that the simulated result matches with the theoretical expression very well, which verifies the correctness of our analysis. Therefore, in the following, we will use the theoretical result to examine some asymptotic properties.

We first compare the error performance of the optimal precoder with a uniform power allocation scheme. Consider the case where $N_t =5~\text{and}~ 50$, $\beta = 2$, and using a 16-QAM constellation. The average SEPs are given for different correlation coefficient $\rho$ in Fig.~\ref{fig:unifpwrload}. Again, we can find that as $|\rho|$ increases, the SEP is becoming significantly worse.  The optimal precoder always leads to better error performance even when $N_t =5$. The gap between the optimal precoder and the uniform power allocation strategy becomes larger when $|\rho|$ increases. The reason is that when $|\rho|$ is very small, $\mathbf \Sigma$ is very close to a diagonal matrix with equal diagonal entries. Then, the optimal precoder will degrade into the uniform power allocation case. However, for general $\mathbf \Sigma$, the performance gap is non-negligible.  To show this phenomenon clearly, the ratio of the SNR of individual sub-channel between optimal precoder and uniform power allocation transmitter are given in Fig.~\ref{fig:precodinggain}. The SNR ratio is a monotonic increasing function of $|\rho|$, which verifies the results in Fig.~\ref{fig:unifpwrload}. Therefore, precoding at the transmitter side can yields better performance over the uniform power allocation strategy. Also, increasing the number of transmitter antennas $N_t$ while keeping $\beta=\frac{N_r}{N_t}$ fixed will decrease the transmitted power of each data stream,  but will increase the diversity order, where the limiting performance can be characterized by using the Sezg\"o's theorem.

On the other hand, to demonstrate the convergence rate in terms of the number of the transmitter antennas, the exact theoretical SEPs and their corresponding limits are depicted versus the number of transmitting antennas in Fig.~\ref{fig:varrho}. Without loss of generality, 16-QAM constellation is adopted and the theoretical SEPs are in solid lines while their limits are denoted by dash lines. Three different correlation matrices are generated according to $\rho$. It can be noticed that as the magnitude of the correlation coefficient $\rho$ decreases, the correlation between the adjacent antennas reduces, and as a consequence, the corresponding SEP reduces substantially.  In Fig.~\ref{fig:varsnr}, the limiting SEPs are also plotted against $N_t$ and different SNRs. It is expected that the SEPs drop as SNR increases. In both the above figures, it can be seen clearly that for a given SNR,  as the array size is scaled up, the theoretical SEPs and the corresponding asymptotic results gradually meet together. The approximation is accurate for moderate and large number of antennas. Note that the mean of SNR for each sub-channel is a decreasing function of $|\rho|$ and an increasing function of system SNR $\eta$.  Hence, either decreasing $|\rho|$ or increasing $\eta$ will eventually increase the mean of SNR for each sub-channel, and thus, result in a lower convergence rate for theoretic SEP approach to its limit expression against $N_t$. This phenomenon is also observed in~\cite{Loyka01} in the scenario of the approximation of channel capacity. In Fig.~\ref{fig:varbeta}, the average SEP curves are plotted against $N_t$ and $\beta=N_r/N_t$. As can be observed that, increasing the number of $\beta$ while letting $N_t$ be a constant will reduce the SEP significantly as predicted in Theorem~\ref{szego:sep}. Actually increasing $\beta$ will improve the array gain, which will result in higher end-to-end channel gain.

In addition, we also show the convergence characteristic of the approximated distribution of individual SNR in each receiver branch for the optimally precoded massive MIMO systems. Both the approximated PDF and the simulated PDF are given in Fig.~\ref{fig:asympspec1}, from which it can be seen clearly that the Gaussian approximation is very accurate even when we only have a very small number of antennas, say, 10 transmitter antennas. Moreover, the case with different correlation coefficients but of the same number of transmitter antennas is studied in Fig.~\ref{fig:asympspec2}. Clearly, as the correlation coefficient $|\rho|$ increases, the mean value of the average SNR decreases as expected and the intervals spanned by the equivalent SNR are also narrowed down.

\section{Conclusions}\label{sec8}
In this paper, we have derived an explicit convex region in terms of the modulated signals, system SNR and channel statistics for the optimal precoder minimizing the average SEP of the ZF detector. A simple expression with a very fast convergence rate  for the SEP limit of the massive MIMO systems with the PAM, PSK and square QAM constellations and the ZF receiver is attained. The intuitive understanding of this convergent process has also been provided in terms of the approximation to the distribution of the individual SNR for each sub-channel.
The main technical approach proposed in this paper to deriving our results is to fully take advantage of the characteristic of the MIMO channels, the structure of the transmitter as well as of the ZF receiver, the Szeg\"o's theorem~\cite{Gray06} on large Hermitian Toeplitz matrices, and the well-known limit:  $\lim_{x\to\infty}(1+1/x)^x=e$.
Numerical results showed that when second-order channel statistics are available to the transmitter, the optimally precoded massive MIMO system outperforms its uniformly precoded counterpart, especially when the antenna correlation is strong.


\section*{Appendix}
\begin{appendices}
\subsection{Proof of Lemma~\ref{lemma:SEP}}\label{appendix:lemmaSEP}
We analyze the SEP expression for $M$-ary PAM, PSK, and QAM constellations as follows:
\subsubsection{PAM signals}
The SEP for $M$-ary PAM signal $s_k$ is $\text{P}_{\text{PAM}}({\mathbf H}, {\mathbf F}, s_k)=\frac{2(M-1)}{M}Q\left(\sqrt{\frac{6\eta }{(M^2-1)[\left({\mathbf F}^H{\mathbf H}^H{\mathbf H}{\mathbf F}\right)^{-1}]_{kk}}}\right)$.
Therefore, the arithmetic average of all SEPs in one block is
\begin{align}\label{avepamsep}
&\text{P}_{\text{PAM}}({\mathbf H}, {\mathbf F})\nonumber\\
&=\frac{2(M-1)}{M N_t}\sum_{k=1}^{N_t}Q\left(\sqrt{\frac{6\eta   }{(M^2-1) [\left({\mathbf F}^H{\mathbf H}^H{\mathbf H}{\mathbf
			F}\right)^{-1}]_{kk}}}\right).
\end{align}
For our purpose, we now prefer to use another expression for Gaussian $Q$-function~\cite{simon1998},
i.e.,
\begin{equation}\label{q}
Q(x)=\frac{1}{\pi}\int_0^{\frac{\pi}{2}}\text{exp}\left(-\frac{x^2}{2\sin^2\theta}\right)\,d\theta,\quad x \ge 0.
\end{equation}
Substituting~(\ref{q}) into~(\ref{avepamsep})
yields
\begin{align}\label{avepamsep1}
&\text{P}_{\text{PAM}}({\mathbf H}, {\mathbf F})= \frac{2(M-1)}{M N_t\pi}\sum_{k=1}^{N_t} \int_0^{\frac{\pi}{2}}...\nonumber\\
&\quad\text{exp}\left(-\frac{3 \eta} {(M^2-1)[\left({\mathbf F}^H{\mathbf H}^H{\mathbf
		H}{\mathbf F}\right)^{-1}]_{kk}\sin^2\theta}\right)\,d\theta.
\end{align}
It is known~\cite[Th.\,3.2.12]{muirhead1982} that
$
\gamma= \frac{[\left({\mathbf F}^H{\mathbf \Sigma}{\mathbf F}\right)^{-1}]_{kk} }{[\left({\mathbf F}^H{\mathbf H}^H {\mathbf H}{\mathbf F}\right)^{-1}]_{kk}}$
is subject to ${\mathcal X}^2_{2(N_r-N_t+1)}$, i.e., its density function is
\begin{equation}\label{eqn:chipdf}
f(\gamma) = \frac{1}{\Gamma(N_r-N_t +1)}e^{-\gamma }\gamma^{N_r-N_t }\quad\textrm{for } \gamma>0,
\end{equation}
where $\Gamma(x)$ denotes the gamma function. Now, taking the
expectation in~(\ref{avepamsep1}) over random channel ${\mathbf H}$
yields
\begin{align}\label{avepamsep2}
&\nonumber\text{P}_{\text{PAM}}({\mathbf F})
\stackrel{(a)}{=} \frac{2(M-1)}{M N_t \pi}\sum_{k=1}^{N_t }\int_0^{\frac{\pi}{2}}...\nonumber\\ &\qquad \mathbb{E}_{{\mathbf H}}\text{exp}\left(-\frac{3\eta } {(M^2-1)[\left({\mathbf F}^H{\mathbf H}^H{\mathbf
		H}{\mathbf F}\right)^{-1}]_{kk}\sin^2\theta}\right)\,d\theta \nonumber \\
&\stackrel{(b)}{=} \frac{2(M-1)}{M N_t \pi}\sum_{k=1}^{N_t}\int_0^{\frac{\pi}{2}}...\nonumber\\ &\qquad\mathbb{E}_{\gamma}\text{exp}\left(-\frac{3 \eta   \gamma } {(M^2-1)[\left({\mathbf F}^H{\mathbf \Sigma }{\mathbf F}\right)^{-1}]_{kk}\sin^2\theta}\right)\,d\theta \nonumber \\
&\stackrel{(c)}{=}  \frac{2(M-1)}{M N_t \pi \Gamma(N_r-N_t +1)}\sum_{k=1}^{N_t} \int_0^{\frac{\pi}{2}} \int_0^\infty\ldots \nonumber \\
&\text{exp}\left[-{\left(1 + \frac{3\eta
	} {(M^2-1)[\left({\mathbf F}^H{\mathbf \Sigma }{\mathbf F}\right)^{-1}]_{k}\sin^2\theta} \right)} \gamma \right] \gamma^{N_r-N_t} d\gamma d\theta \nonumber\\
&\stackrel{(d)}{=} \frac{2(M-1)}{M N_t \pi}\sum_{k=1}^{N_t}\int_0^{\frac{\pi}{2}}...\nonumber\\
&\quad\left(1+\frac{3 \eta  }{(M^2-1)[\left({\mathbf F}^H{\mathbf \Sigma}{\mathbf
		F}\right)^{-1}]_{kk}\sin^2\theta}\right)^{-(N_r-N_t+1)}d\theta\nonumber\\
&=\frac{1}{N_t}\sum_{k=1}^{N_t}G_{\text{PAM}}\left([\left({\mathbf
	F}^H{\mathbf \Sigma}{\mathbf F}\right)^{-1}]_{kk}\right),
\end{align}
where $(a)$ is obtained by inserting \eqref{q} into \eqref{avepamsep} and then averaging over the channel statistics, (b) is true for the definition of $\gamma$ above, (c) holds due to~\eqref{eqn:chipdf}, and $(d)$  can be attained by finishing the integral over the distribution of $\gamma$ given above. In addition, function $G_{\text{PAM}}(x)$ is defined as
$G_{\text{PAM}}(x)= \frac{2(M-1)}{M \pi} \int_0^{\pi/2}\left(1+\frac{3\eta   }{(M^2-1)x\sin^2\theta}\right)^{-(N_r-N_t+1)}d\theta$.
The second-order derivative of $G_{\text{PAM}}(x)$ is given by
\begin{align*}
&\frac{d^2G_{\text{PAM}}(x)}{dx^2}\\
&=\frac{2(M-1)}{M\pi}\int_0^{\frac{\pi}{2}}\!\!\left(1+\frac{3 \eta  }{(M^2-1)  x\sin^2\theta}\right)^{-(N_r-N_t+3)} \nonumber\\
&\quad\times \frac{3  \eta (N_r-N_t+1) }{(M^2-1) x^4 \sin^2\theta }\left(\frac{ 3 \eta (N_r-N_t) }{(M^2 -1) \sin^2\theta} -2t \right) d\theta.
\end{align*}

Since ${ 3 \eta (N_r-N_t) } /{\left((M^2 -1) \sin^2\theta\right)} -2t \ge{ 3 \eta (N_r-N_t) }/{\left((M^2 -1)  \right)} -2t$, then if the following condition is satisfied, i.e.,
\begin{align*}
0<x\le \frac{ 3\eta  (N_r-N_t) }{2(M^2 -1) } =T_{\rm PAM},
\end{align*}
we have ${d^2G_{\text{PAM}}(x)}/{dx^2} \ge 0$.  This implies that $G_{\text{PAM}}(x)$ is convex in this interval.

\subsubsection{PSK signals}
The SEP for $M$-ary PSK signal $s_k$ is
\begin{align}\label{psksep}
&\text{P}_{\text{PSK}}({\mathbf H}, {\mathbf F},
s_k)=\frac{1}{\pi}\int_0^{\frac{(M-1)\pi}{M}}...\nonumber\\
&\qquad\qquad \exp\left(-\frac{\eta\sin^2(\pi/M)}{[\left({\mathbf F}^H{\mathbf H}^H{\mathbf H}{\mathbf F}\right)^{-1}]_{kk}  \sin^2 \theta}\right)\,d\theta.
\end{align}
Therefore, the arithmetic mean of all SEPs is
\begin{align}\label{avepsksep1}
&\text{P}_{\text{PSK}}({\mathbf H}, {\mathbf F})=
\frac{1}{N_t \pi}\sum_{k=1}^{N_t}\int_0^{\frac{(M-1)\pi}{M}}\nonumber\\
&\qquad\text{exp} \left(-\frac{ \eta \sin^2(\pi/M)}{[\left({\mathbf F}^H{\mathbf H}^H{\mathbf H}{\mathbf F}\right)^{-1}]_{kk} \sin^2 \theta}\right)\,d\theta.
\end{align}
Similarly, taking the expectation of~(\ref{avepsksep1}) over random channel ${\mathbf H}$
produces
\begin{align*}\label{avepsksep2}
&\nonumber \text{P}_{\text{PSK}}({\mathbf F})=\frac{1}{N_t \pi}\sum_{k=1}^{N_t}\int_0^{\frac{(M-1)\pi}{M}} ...\\
&\qquad\qquad \left(1+\frac{ \eta \sin^2(\pi/M)}{[\left({\mathbf F}^H{\mathbf \Sigma}{\mathbf F}\right)^{-1}]_{kk}\sin^2\theta}\right)^{-(N_r-N_t+1)}d\theta\\
&=\frac{1}{N_t}\sum_{k=1}^{N_t}G_{\text{PSK}}\left([\left({\mathbf F}^H{\mathbf \Sigma}{\mathbf F}\right)^{-1}]_{kk}\right),
\end{align*}
where function $G_{\text{PSK}}(x)$ is defined as
\begin{equation*}\label{gpsk}
G_{\text{PSK}}(x)=\frac{1}{\pi}\int_0^{\frac{(M-1)\pi}{M}} \left(1+\frac{ \eta\sin^2(\pi/M)}{ x\sin^2\theta}\right)^{-(N_r-N_t+1)}d\theta.
\end{equation*}
Now, the second-order derivative of $G_{\text{PSK}}(x)$ is
\begin{align*}
&\frac{d^2G_{\text{PSK}}(x)}{dx^2}\!=\!\frac{1}{ \pi}\int_0^{ \frac{(M-1)\pi}{M}}\!\!\left(1+\frac{ \eta \sin^2{(\pi/M)}}{ x\sin^2\theta}\right)^{-(N_r-N_t+3)}\\
&\!\!\times\!\!\frac{\eta (N_r-N_t+1)  \sin^2{(\frac{\pi}{M})}}{ x^4 \sin^2\theta }\!\left(\frac{ \eta (N_r-N_t) \sin^2{(\frac{\pi}{M})}}{ \sin^2\theta }-2t\!\right) d\theta.
\end{align*}
Since $\eta (N_r-N_t) \sin^2{(\pi/M) }/( \sin^2\theta)-2t\ge \eta (N_r-N_t) \sin^2{(\pi/M) }-2t\ge 0$, we have that if
\begin{align}\label{pskcondt}
0<x\le \frac{\eta(N_r-N_t) \sin^2{(\pi/M)}}{2}=T_{\rm PSK},
\end{align}
then, in this interval, $d^2 G_{\text{PSK}}(x)/dx^2\ge 0$.
This shows that $G_{\text{PSK}}(x)$ is convex in this range.
\subsubsection{QAM signals}
The SEP for $M$-ary QAM signal $s_k$ is
\begin{align}\label{qamsep}
&\nonumber \text{P}_{\text{QAM}}({\mathbf H}, {\mathbf F},
s_k)\\
&=4\left(1-1/ \sqrt{M} \right)Q\bigg(\sqrt{\frac{3 \eta}{(M-1)[\left({\mathbf F}^H{\mathbf H}^H{\mathbf H}{\mathbf F}\right)^{-1}]_{kk}}}\bigg)\nonumber\\
&-4\left(1-1/\sqrt{M} \right)^2Q^2\bigg(\sqrt{\frac{3 \eta}{(M-1)[\left({\mathbf F}^H{\mathbf H}^H{\mathbf H}{\mathbf F}\right)^{-1}]_{kk}}}\bigg).
\end{align}
The first term in~(\ref{qamsep}) can be replaced by~(\ref{q}). Similarly, $Q^2(\cdot)$ function also has a very nice
formula~\cite{simon1998},
\begin{equation}\label{qq}
Q^2(x)=\frac{1}{\pi}\int_0^{\pi/4}\text{exp}\left(-\frac{x^2}{2\sin^2\theta}\right)\,d\theta.
\end{equation}
Substituting~(\ref{q}) and~(\ref{qq})
into~(\ref{qamsep}) and then, taking the expectation over the random channel
matrix, we can obtain
\begin{align}
&\text{P}_{\text{QAM}}({\mathbf F})
=\frac{1}{N_t}\sum_{k=1}^{N_t}G_{\text{QAM}}\left([\left({\mathbf F}^H{\mathbf \Sigma}{\mathbf F}\right)^{-1}]_{kk}\right)\label{eqn:sepqam},
\end{align}
where function $G_{\text{QAM}}(x)$ is defined as
\begin{align}\label{gqam}
&G_{\text{QAM}}(x)\nonumber\\
&=\frac{4(\sqrt{M}-1)} {\sqrt{M} \pi}\int_{\frac{\pi}{4}}^{\frac{\pi}{2}}\left(1+\frac{3\eta }{2(M-1) x \sin^2\theta}\right)^{-(N_r-N_t+1)}\!\!\!\!d\theta \nonumber\\
& + \frac{4(\sqrt{M}-1)} {M\pi}\int_{0}^{\pi/4} \left(1+\frac{3 \eta }{2(M-1) x \sin^2\theta}\right)^{-(N_r-N_t+1)}\!\!\!\!\!\!d\theta.
\end{align}	
For QAM signals, the second-order derivative of $G_{\text{QAM}}(x)$ is
\begin{align*}
&\frac{d^2G_{\text{QAM}}(x)}{dx^2}=
\Big(\frac{4}{\pi}-\frac{4}{\sqrt{M}\pi}\Big)\\
&\times \int_{\frac{\pi}{4}}^{\frac{\pi}{2}}\Big(1+ \frac{3\eta }{2(M-1)x\sin^2\theta}\Big)^{-(N_r-N_t+3)}\\
&\times \frac{3\eta (N_r-N_t+1) }{ 2(M-1)  x^4 \sin^2\theta} \left(\frac{3\eta(N_r-N_t ) }{2(M-1) \sin^2\theta}-2t \right) d\theta\\
&+ 4\frac{(1-1/\sqrt{M})}{(\sqrt{M}\pi}\int_0^{\frac{\pi}{4}}\left(1+ \frac{3 \eta}{2(M-1) x\sin^2\theta}\right)^{-(N_r-N_t+3)}\\
&\times \frac{3 \eta (N_r-N_t+1) }{ 2(M-1) x^4 \sin^2\theta } \left(\frac{3\eta (N_r-N_t ) }{2(M-1) \sin^2\theta}-2t \right) d\theta.
\end{align*}
Since $3\eta(N_r-N_t) /(2(M-1) \sin^2\theta)-2t \ge
3\eta(N_r-N_t) /(2(M-1))-2t \ge 0$, if the following
condition meets,
\begin{eqnarray}\label{qamcondt}
0<x\le \frac{3\eta (N_r-N_t) }{4(M-1)}=T_{\rm QAM},
\end{eqnarray}
then, $d^2 G_{\text{QAM}}(x)/dx^2\ge 0$
and as a result, $G_{\text{QAM}}(x)$ is a convex function in this
interval.

Combing all the above results, we complete the proof of Lemma~\ref{lemma:SEP}. 

\subsection{Proof of Proposition~\ref{prop:convexity}}\label{appendix:convexity}
To develop an explicit constraint from~\eqref{snr}, we need to introduce the following two lemmas.
\begin{lemma}[Rayleigh-Ritz]\cite{horn1985}\label{lem:ray}
	Let ${\mathbf A} \in {\mathbb C}^{K\times K}$ be Hermitian and let $\zeta_1({\mathbf A}) \le \zeta_2({\mathbf A}) \le \ldots \le \zeta_K({\mathbf A})$ be the eigenvalues of ${\mathbf A}$.
	Then
	\begin{align*}
	&\zeta_1({\mathbf A}) = \min_{ \left\|{\mathbf x}\right\|\neq 0 } \frac{{\mathbf x}^H {\mathbf A} {\mathbf x} }{{\mathbf x}^H {\mathbf x}},\qquad
	\zeta_K({\mathbf A}) = \max_{ \left\|{\mathbf x}\right\|\neq 0 } \frac{{\mathbf x}^H {\mathbf A} {\mathbf x} }{{\mathbf x}^H {\mathbf x}}.
	\end{align*} 
\end{lemma}
Since $({\mathbf F^H}{\mathbf \Sigma}{\mathbf F})^{-1}$ is Hermitian and $ [({\mathbf F^H}{\mathbf \Sigma}{\mathbf F})^{-1}]_{kk} = {\mathbf e}_k^H  ({\mathbf F^H}{\mathbf \Sigma}{\mathbf F})^{-1} {\mathbf e}_k$, where ${\mathbf e}_k =[0, \cdots, 0,1,0, \cdots,0]^H$ has 1 only in the $k$-th entry, by Lemma~\ref{lem:ray} we have
\begin{align*}
\zeta_1\big( ({\mathbf F^H}{\mathbf \Sigma}{\mathbf F})^{-1}\big)  \le [({\mathbf F^H}{\mathbf \Sigma}{\mathbf F})^{-1}]_{kk} \le \zeta_{N_t}\big(({\mathbf F^H}{\mathbf \Sigma}{\mathbf F})^{-1}\big),~\forall k,
\end{align*}
where $\zeta_1\big( ({\mathbf F^H}{\mathbf \Sigma}{\mathbf F})^{-1}\big) $ and $\zeta_{N_t}\big(({\mathbf F^H}{\mathbf \Sigma}{\mathbf F})^{-1}\big)$ are the minimum and maximum eigenvalues of $ ({\mathbf F^H}{\mathbf \Sigma}{\mathbf F})^{-1}$, respectively and the equality is attainable~when ${\mathbf F}$ diagonalizes $\mathbf \Sigma$. Therefore, if ${\mathbf F}$ satisfies
\begin{align}\label{eqn:newconstraint}
\zeta_{N_t}\big( ({\mathbf F^H}{\mathbf \Sigma}{\mathbf F})^{-1} \big)\le T,
\end{align}
then, such ${\mathbf F}$ also satisfies~\eqref{snr}. To further simplify the constraint, we need another lemma.
\begin{lemma}[Ostrowski]\cite{horn1985}\label{Ostrowski2} Let ${\mathbf A}\in {\mathbb C}^{K\times K}$ be Hermitian and ${\mathbf S} \in {\mathbb C}^{K\times K}$ be nonsingular. If we let the eigenvalues of $\mathbf A$ and ${\mathbf S} {\mathbf S}^H$ be given by $\zeta_1({\mathbf A})\le \zeta_2({\mathbf A})\le \ldots\le \zeta_K({\mathbf A})$ and $\zeta_1({\mathbf S} {\mathbf S}^H)\le \zeta_2({\mathbf S} {\mathbf S}^H)\le \ldots\le \zeta_K({\mathbf S} {\mathbf S}^H)$, respectively, then, for each $i=1,2,\ldots, K$, there exists a positive real number $\kappa_i$ such that $\zeta_1({\mathbf S} {\mathbf S}^H) \le \kappa_i\le \zeta_K({\mathbf S} {\mathbf S}^H)$ and $\zeta_i({\mathbf S} {\mathbf A}{\mathbf S}^H) =\kappa_i \zeta_i({\mathbf A})$. 
\end{lemma}

Let the eigenvalue decomposition (EVD) of ${\mathbf\Sigma}$ and the singular
value decomposition (SVD) of ${\mathbf F}$ be
\begin{subequations}\label{eqn:defsigmaprecoding}
\begin{align}
{\mathbf\Sigma}&={\mathbf W}{\mathbf\Lambda}{\mathbf W}^H,\label{eqn:defsigma}\\
{\mathbf F}&= {\mathbf U}{\mathbf D}{\mathbf V},\label{eqn:defprecoding}
\end{align}
\end{subequations}
where
${\mathbf W}$, ${\mathbf U}$, and ${\mathbf V}$ are all unitary matrices. We also assume ${\mathbf\Lambda}={\rm
	diag}(\lambda_1, \lambda_2, \cdots, \lambda_{N_t})$, where
$0<\lambda_1\le \lambda_2 \le \cdots \le \lambda_{N_t}$, and ${\mathbf D}={\rm
	diag}(\sqrt{d_1}, \sqrt{d_2}, \cdots, \sqrt{d_{N_t}})$, where $0 < d_1\le d_2 \le \dots \le d_{N_t}$, since $\mathbf F$ is assumed to be of full-rank (nonsingular).
Then, by Lemma~\ref{Ostrowski2} with $K=N_t$, ${\mathbf S} ={\mathbf F}^{-1}$ and ${\mathbf A} ={\mathbf \Sigma}^{-1}$, we have
\begin{align}
&\zeta_{N_t}\big(( {\mathbf F^H}{\mathbf \Sigma}{\mathbf F})^{-1} \big)
=\zeta_{N_t}\big( {\mathbf F^{-1}}{\mathbf \Sigma}^{-1}{\mathbf F}^{-H} \big)\nonumber\\
&\le \zeta_{N_t}({\mathbf \Sigma}^{-1}) \zeta_{N_t}\big(({\mathbf F}^H {\mathbf F})^{-1}\big)
=\frac{1}{\lambda_1 d_1},
\end{align}
where the equality is also attainable when ${\mathbf U} ={\mathbf W}$. As $d_1=\zeta_1( {\mathbf F} ^H {\mathbf F})$, then if
\begin{align}\label{newcvxconstraint}
\zeta_1( {\mathbf F} ^H {\mathbf F})=d_1 \ge \frac{1}{\lambda_{1} T  },
\end{align}
we can conclude that ${\mathbf F}$ satisfies constraint~\eqref{snr}. This completes the proof of Proposition~\ref{prop:convexity}.

\subsection{Proof of Theorem~\ref{th:opt-zf}}\label{append:theoremoptzf}
First, by Lemma~\ref{lemma:SEP}, the average SEP with precoding matrix $\mathbf{F}$ is given by
\begin{align}
\text{P}({\mathbf F})
=\frac{1}{N_t}\sum_{k=1}^{N_t}G\left([\left({\mathbf F}^H{\mathbf \Sigma}{\mathbf F}\right)^{-1}]_{kk}\right),
\end{align}
where $G(x)$ is convex for $0<x<T$. Now, following the same way as~\cite{ding03, Palomar03, kiessling04} and applying the Jensen's inequality~\cite{cover91} to function $G(x)$ under the constraint~\eqref{newcvxconstraint} result in
\begin{eqnarray}\label{thresh2}
\frac{1}{N_t}\sum_{k=1, d_1 \ge \frac{1}{\lambda_1 T} }^{N_t} G\left([({\mathbf
	F^H}{\mathbf\Sigma}{\mathbf F})^{-1}]_{kk}\right)\nonumber\\
\ge
G\left(\frac{1}{N_t}\sum_{k=1}^{N_t} [({\mathbf
	F^H}{\mathbf\Sigma}{\mathbf F})^{-1}]_{kk}\right),
\end{eqnarray}
where the equality in~(\ref{thresh2}) holds if and only if
$[({\mathbf F^H}{\mathbf\Sigma}{\mathbf F})^{-1}]_{11}=[({\mathbf
	F^H}{\mathbf\Sigma}{\mathbf F})^{-1}]_{22}=\cdots=[({\mathbf
	F^H}{\mathbf\Sigma}{\mathbf F})^{-1}]_{N_t N_t}$.
Recall that, in~\eqref{eqn:defsigmaprecoding}, we let
${\mathbf\Sigma}={\mathbf W}{\mathbf\Lambda}{\mathbf W}^H$, and ${\mathbf F}= {\mathbf U}{\mathbf D}{\mathbf V}$, then by a well known trace-inequality~\cite{Marshall79}, we have
\begin{align}
&{\rm tr}\left(({\mathbf F^H}{\mathbf\Sigma}{\mathbf F})^{-1}\right)
={\rm tr}\left( {\mathbf F}^{-1}  {\mathbf \Sigma }^{-1} {\mathbf F}^{-H}   \right)={\rm tr}\left( (  {\mathbf F}  {\mathbf F}^H)^{-1} {\mathbf \Sigma }^{-1}    \right) \nonumber\\
&={\rm tr}\left( {\mathbf U} {\mathbf D}^{-2} {\mathbf U}^H {\mathbf W}{\mathbf\Lambda}^{-1}{\mathbf W}^H  \right)
\ge \sum_{k=1}^{N_t}d^{-1}_{N_t +1-k} \lambda_{k}^{-1}\label{eqn:lowerbound},
\end{align}
where the equality in~(\ref{eqn:lowerbound}) holds if ${\mathbf
U}={\mathbf W}{\mathbf P}$, in which  $\mathbf P$ is an anti-diagonal permutation matrix given by
$$
{\mathbf P} = \begin{bmatrix}
0&  \cdots & 0 & 1\\
0&\cdots& 1&0\\
\vdots&\ddots&\vdots&\vdots\\
1&\cdots&0&0
\end{bmatrix}.
$$
Then, using the Cauchy-Schwarz inequality, we can attain
$\sum_{k=1}^{N_t} \sqrt{d_{N_t+1-k}} \cdot \frac{ 1 }{\sqrt{d_{N_t+1-k}  \lambda_{k} } }\le
\sqrt{\sum_{\ell=1}^{N_t} d_\ell } \cdot \sqrt{ \sum_{k=1}^{N_t} \frac{ 1 }{d_{N_t+1-k}  \lambda_{k}}}$.
Combining this with the power constraint ${\rm tr}({\mathbf F}^H{\mathbf F})=1$ gives us
\begin{eqnarray}\label{tr1}
\sum_{k=1}^{N_t}d^{-1}_{N_t +1 -k}\lambda_{k}^{-1} \ge \left(\sum_{k=1}^{N_t }  \lambda_{k}^{-1/2}    \right)^2.
\end{eqnarray}
The equality in~(\ref{tr1}) holds if and only if
\begin{eqnarray}\label{trcondt}
d_{N_t+1-k}=\frac{ \lambda_{k}^{-1/2} }{\sum_{\ell=1}^{N_t} \lambda_{\ell}^{-1/2}}, \quad k=1,2, \ldots, N_t.
\end{eqnarray}
Since $G(x)$ monotonically increases, combining~\eqref{trcondt} with~\eqref{thresh2} leads to
\begin{align*}
\frac{1}{N_t}\sum_{k=1, d_1 \ge \frac{1}{\lambda_1 T} }^{N_t} G\left([({\mathbf
	F^H}{\mathbf\Sigma}{\mathbf F})^{-1}]_{kk}\right)\ge
G\bigg({\Big(\sum_{k=1}^{N_t}
	\lambda_k^{-1/2} \Big)^2}/{N_t} \bigg),
\end{align*}
where the equality holds if ${\mathbf U}={\mathbf W}{\mathbf P}$, the square of the singularvalues of ${\mathbf F}$ meets~\eqref{trcondt} and ${\mathbf V}$ is chosen as the normalized DFT matrix.
Therefore, the optimal solution is 
\begin{eqnarray}\label{formu2solu}
\widetilde{\mathbf F}_1=\frac{1}{\sqrt{\rm tr({\mathbf \Lambda}^{-1/2})}}{\mathbf W}{\mathbf \Lambda}^{-1/4} \widetilde{\mathbf V}_1,
\end{eqnarray}
where $\widetilde{\mathbf V}_1$ is an arbitrary unitary matrix. By specifically chose $\widetilde{\mathbf V}_1$ be a normalized unitary maxtrix, we complete the proof of Theorem~\ref{th:opt-zf}. 

\subsection{Proof of Theorem~\ref{szego:sep}}\label{append:szegosep}
 By using Lemma~\ref{thm:szeg} with $K=N_t, {\mathbf T}_{N_t}={\mathbf\Sigma}$ and ${\mathtt F}(x)=1/\sqrt{x}$, we have
\begin{align}\label{eqn:szegolimit}
	\Lambda=\lim_{N_t \to \infty} \frac{\sum_{k=1}^{N_t} \lambda_k^{-1/2}}{N_t} = \frac{1}{2\pi} \int_{0}^{2\pi} \frac{1}{\sqrt{s_{\Sigma}(\omega)}} d\omega
\end{align}
where $\lambda_1, \lambda_2 \ldots, \lambda_{N_t}$ are the eigenvalues of $\mathbf \Sigma$.
	For notational simplicity, let $\bar{\lambda}_{N_t}={\big(\sum_{k=1}^{N_t}
		\lambda_k^{-1/2} \big)}/{N_t}$. Now, using the optimal precoder given in Theorem~\ref{th:opt-zf}, the resulting minimum average SEP is
	$\lim_{N_t \to \infty} \text{P}_{\rm min}(\widetilde{\mathbf F})
	=   \lim_{N_t \to \infty}  G\Big({\big(\sum_{k=1}^{N_t}
		\lambda_k^{-1/2}\big)^2}/{N_t}\Big)$.
	Correspondingly, for PAM signal, we obtain
	\begin{align}
	&\bar{{\rm P}}_{\rm opt, PAM}
	 \mathop{=}^{(a)} \frac{2(M-1)}{M \pi} \nonumber\\
	&~\times\int_0^{\frac{\pi}{2}} \lim_{N_t \to \infty} \Big(1+\frac{3 \eta  }{(M^2-1) N_t \bar{\lambda}_{N_t}^2\sin^2\theta} \Big)^{-(\beta-1) N_t -1}d\theta \nonumber\\
	& \mathop{=}^{(b)}  \frac{2(M-1)}{M \pi} \int_0^{\frac{\pi}{2}} \exp \Big( - \frac{3\eta (\beta-1)}{(M^2-1) \Lambda^2\sin^2\theta} \Big) d\theta \nonumber\\
	&= \frac{2(M-1)}{M} Q\bigg( \sqrt{  \frac{ 6\eta (\beta-1) }{(M^2-1) \Lambda^2 }} \bigg),\label{eqn:pamlimit}
	\end{align}
	where equality $(a)$ follows from the fact that $\left( 1+\frac{3 \eta  }{(M^2-1) N_t \bar{\lambda}_{N_t}^2\sin^2\theta} \right)^{-(\beta-1) N_t-1} < 1$ for all $N_t \ge 1$ and $\theta \in [0, \pi/2]$
	and thus, by the Lebesgue's Dominated Convergence Theorem~\cite{Bartle95}, we can change the order of limitation and integration. The equality $(b)$ is due to the well-known limit of the Euler's number. Following the similar argument,
	for PSK signal, we have
	\begin{align}
	&\bar{{\rm P}}_{\rm opt, PSK}= \frac{1}{\pi}\int_0^{\frac{(M-1)\pi}{M}}\!\!\!\!\!\lim_{N_t \to \infty} \left(1+\frac{\eta \sin^2(\pi/M)}{ N_t  \bar{\lambda}_{N_t}^2\sin^2\theta}\right)^{-(\beta-1)N_t -1}\!\!\!\!\!\!\!\!\! d\theta \nonumber\\
	&=\frac{1}{\pi} \int_0^{\frac{(M-1)\pi}{M}} \exp \left( - \frac{ \eta (\beta-1) \sin^2(\pi/M)}{\Lambda^2\sin^2\theta} \right) d\theta\label{eqn:psklimit}
	\end{align}
	and for QAM signal, we can attain
	\begin{align}
	&\nonumber\bar{{\rm P}}_{\rm opt, QAM}\nonumber\\
	&=\frac{4(\sqrt{M}-1)} {\sqrt{M}\pi}\int_0^{\frac{\pi}{2}}...\nonumber\\
	&\quad\lim_{N_t \to \infty}  \left(1+\frac{3\eta }{2(M-1) N_t \bar{\lambda}_{N_t}^2 \sin^2\theta}\right)^{-(\beta-1)N_t -1}d\theta\nonumber\\
	&- \frac{4(\sqrt{M}-1)^2} {M \pi}\int_{0}^{\frac{\pi}{4}}...\nonumber\\
	&\quad\lim_{N_t \to \infty}  \left(1+\frac{3 \eta}{2(M-1)N_t \bar{\lambda}_{N_t}^2 \sin^2\theta}\right)^{-(\beta-1)N_t -1}d\theta\nonumber\\
	&= \frac{4(\sqrt{M}-1)}{\sqrt{M}} Q\bigg(\sqrt{\frac{3\eta(\beta-1) }{(M-1) \Lambda^2}} \bigg)\nonumber\\
	&\quad-\frac{4(\sqrt{M}-1)^2}{M} Q^2\bigg( \sqrt{ \frac{3\eta (\beta-1) }{(M-1)\Lambda^2}} \bigg).\label{eqn:qamlimit}
	\end{align}
	This completes the proof of Theorem~\ref{szego:sep}.
\subsection{Proof of Corollary~\ref{corollary:kms}}\label{append:kms}
We know that, $s_{\Sigma}(\omega) = \sum_{k=-\infty}^{\infty} \sigma(k) e^{-jk\omega}
= \frac{1-|\rho|^2}{ 1+|\rho|^2 -2 \mathtt{Re}[\rho e^{j\omega}] }.$
for $ 0 < |\rho| < 1$, we have $s_{\Sigma}(\omega)\ge\frac{1-|\rho|}{ 1+|\rho|}$ and as a result, $L_{s_{\Sigma}}\ge \frac{1-|\rho|}{ 1+|\rho|}>0$.
Now by Lemma~\ref{thm:szeg} with ${\mathtt F}(x)=1/\sqrt{x}$, we attain
\begin{align*}
\Lambda_{KMS} 
&= \frac{1}{\pi\sqrt{1-|\rho|^2}} \int_{0}^{\pi} \sqrt{1+|\rho|^2 -2|\rho| \cos \omega}  d\omega\\
&= \frac{2}{\pi}\sqrt{ \frac{1+|\rho|}{1-|\rho|}} {\mathtt E} \left( \frac{ 2 \sqrt{|\rho|}}{ 1+|\rho|} \right).
\end{align*}
Combining this with Theorem~\ref{szego:sep} completes the proof of Corollary~\ref{corollary:kms}.

\subsection{Proof of Theorem~\ref{thm:uniformpallo}}\label{appendix:thmkms}
Note that KMS matrix ${\bf \Sigma }$ has a simple tridiagonal inverse~\cite{Dow2003}, given by
\begin{align*}
	{\mathbf \Sigma}^{-1} =\frac{1}{1-|\rho|^2} \begin{bmatrix}
		1 & -\rho & 0 &\cdots& 0\\
		-\rho^* & 1+ |\rho|^2 & -\rho &\cdots& 0\\
		&\ddots&\ddots &\ddots&  \\
		0 & \cdots & -\rho^* & 1+|\rho|^2 & -\rho \\
		0 & \cdots & 0 & -\rho^* & 1
	\end{bmatrix}.
\end{align*}
Combining this with~\eqref{eqn:sepqam} and the uniform precoder yields
$\text{P}_{\text{QAM}}({\bar{\mathbf F}})
=\frac{2}{N_t}{G}_{\rm QAM}\Big(\frac{N_t}{1-|\rho|^2} \Big) + \frac{N_t-2}{N_t}{G}_{\rm QAM}\Big( \frac{N_t (1+|\rho|^2)}{1-|\rho|^2}   \Big)$.
Since $\lim_{N_t \to \infty} \left(1+\frac{3\eta (1-|\rho|^2) }{2(M-1) N_t \Phi \sin^2\theta}\right)^{-(N_r-N_t+1)}=\exp\bigg({ -\frac{3 \eta(\beta-1)(1-|\rho|^2) }{2(M-1) \Phi \sin^2\theta}}\bigg)$,
where $\Phi=1$ or $1+|\rho|^2$, we have
\begin{align*}
&\bar{{\rm P}}_{\rm{U, QAM}} = \lim_{N_t \to \infty} {G}_{\rm QAM}\Big( \frac{N_t (1+|\rho|^2)}{1-|\rho|^2}   \Big) \nonumber\\
&=\frac{4(\sqrt{M}-1)}{\sqrt{M}}  Q\left(\sqrt{\frac{3\eta(\beta-1) (1-|\rho|^2) }{(M-1)(1+|\rho|^2)  }} \right)\nonumber\\ &\quad-\frac{4(\sqrt{M}-1)^2}{M}  Q^2\left( \sqrt{ \frac{3\eta(\beta-1)(1-|\rho|^2) }{(M-1) (1+|\rho|^2)   }} \right).
\end{align*}
The expressions of $\bar{{\rm P}}_{{\rm U, PAM}}$ and $\bar{{\rm P}}_{{\rm U, PSK}}$ can be obtained in a similar fasion and hence are omitted for brevity. This completes the proof of Theorem~\ref{thm:uniformpallo}.
\end{appendices}


\small

\bibliographystyle{ieeetr}

\bibliography{tzzt}

\end{document}